\documentclass[nofootinbib,
 reprint,
 amsmath,amssymb,
 aps,
]{revtex4-2}

\usepackage{graphicx}
\usepackage{dcolumn}
\usepackage{bm}

 \usepackage{amsmath}
\usepackage{epstopdf}
\usepackage{float}
\usepackage{hyperref}
\usepackage{enumitem}
\usepackage[caption=false]{subfig}
\usepackage{color}
\usepackage[T1]{fontenc}
\usepackage[utf8]{inputenc}
\usepackage[toc,page]{appendix}
\usepackage[usenames,dvipsnames]{xcolor}
\usepackage[normalem]{ulem}
\usepackage{lipsum, babel}

\usepackage{physics}
\usepackage{graphicx}
\usepackage{cleveref}

\newcommand{\lr}[3]{\left#1 #2\right#3}

\newcommand{\be}{\begin{equation}}
\newcommand{\ee}{\end{equation}}
\newcommand{\ba}{\begin{eqnarray}}
\newcommand{\ea}{\end{eqnarray}}

\newcommand{\beq}{\begin{equation}}
\newcommand{\eeq}{\end{equation}}
\newcommand{\beqa}{\begin{eqnarray}}
\newcommand{\eeqa}{\end{eqnarray}}

\begin{document}

\preprint{APS/123-QED}
\title{Optical 
properties of black holes in regularized Maxwell theory}


\author{Tomáš Hale}%
\email{tomas.hale@utf.mff.cuni.cz}
\author{David Kubizňák}%
\email{david.kubiznak@matfyz.cuni.cz}
\author{Jana Menšíková}
\email{Jana.Mensikova@matfyz.cuni.cz}
\affiliation{%
 Institute of Theoretical Physics, Faculty of Mathematics and Physics,
Charles University, V Holešovičkách 2, 180 00 Prague 8, Czech Republic
}%

\date{January 29, 2024}

\begin{abstract}
Regularized Maxwell electrodynamics is a recently discovered theory of non-linear electrodynamics, with a `minimally regularized' field strength of a point charge, that is `very close' to the Maxwell theory in many aspects. In this paper we investigate some of the optical properties of its black holes. Namely, we study geodesics, gravitational red-shift, black hole shadow, as well as investigate the relationship between the behavior of (null geodesic) Lyapunov exponents and the existence of thermodynamic critical points in both canonical and grand-canonical ensembles. \end{abstract}

\maketitle


\section{introduction}
The first (fully covariant) theories of {\em Non-Linear Electrodynamics} (NLE)  were formulated 
in the 1930s as a classical attempt at regularizing the field of a point charge in Maxwell's theory \cite{born1933modified, born1934foundations, born1934quantization, HInfeld}. The key idea was to modify the corresponding electromagnetic Lagrangian, making it sufficiently non-linear, so that the point charge could be characterized by a finite field strength and finite self-energy. Since then, many other theories with various characteristics 
were constructed  -- nowadays NLE provides a general framework for  studying classical electrodynamics with `improved properties', while maintaining the second-order equations of motion and covariant action, see \cite{Sorokin:2021tge} for  a recent review.

Among many NLEs, 
perhaps the most prolific to date is the 
{\em Born--Infeld theory}, developed by M. Born and L. Infeld in \cite{born1934foundations, born1934quantization}. This theory yields a finite point
charge self-energy, possesses an electromagnetic duality \cite{Gibbons:1995cv}, and approaches Maxwell's electrodynamics in the weak field limit. 
It finds applications in string theory \cite{Fradkin:1985qd}, D-brane physics \cite{Leigh:1989jq, Callan:1997kz}, or cosmology \cite{Alishahiha:2004eh}.  
Moreover, as found by Pleba\'nsky \cite{Plebanski:1970zz} (see also \cite{Russo:2022qvz, Mezincescu:2023zny}), it is the unique NLE (apart from Maxwell), in which light rays propagate without {\em birefringence}, that is, the two electromagnetic modes in vacuum have equal phase velocities. Other NLEs thus suffer from birefringence, which can in some cases lead to pathologies, such as \lq superliminal photons\rq, providing a basis for excluding the corresponding NLE. As we shall review in this paper, a common approach to deal with the birefringence is to formulate an effective geometry, which governs the propagation of the corresponding mode in the geometric optics approximation, e.g. \cite{Novello:1999pg}.

The current interest in NLEs, however, goes well beyond the original Born--Infeld theory. For example, it was shown, that the framework of NLEs can naturally provide  `physical sources' for regular black hole spacetimes. In this spirit, the famous regular Bardeen black hole \cite{bardeen1968non} has been identified with a magnetically charged black hole in due NLE \cite{Bardeen}, see also \cite{Balart:2014cga, Fan:2016hvf, Bokulic:2023afx} for other regular black hole models interpreted as solutions of non-linear electrodynamics.
Another example of a very interesting NLE is the so called {\em ModMax theory}  \cite{ModMax, Kosyakov:2020wxv}, which is the most general NLE that admits the same symmetries as Maxwell's theory, namely, the conformal invariance and the electromagnetic duality (see \cite{Russo:2024llm} for a recent discussion of causal NLEs with electromagnetic duality).

In this paper, we focus on yet another recently formulated theory of NLE, called the {\em Regularized Maxwell} (RegMax) theory \cite{newLagrangian}. This theory is in many aspects `very close' to the linear Maxwell electrodynamics. It gives rise to  a minimally regularized field strength of a point charge, replacing the $r^{-2}$ behavior with $(r+r_0)^{-2}$, for some positive constant $r_0$. 
At the same time, many of its self-gravitating solutions are remarkably `Maxwell-like'. Namely, it is the unique NLE theory, whose radiative solutions can be (similar to Maxwell) found in the Robinson--Trautman class \cite{newLagrangian} and are (contrary to Maxwell) well posed \cite{Tahamtan:2023tci}. Moreover, 
it is the unique NLE apart from Maxwell whose slowly rotating charged solutions are fully characterized by the electrostatic potential \cite{slowlyrotating}. 
A general overview of the basic properties of RegMax theory, including black hole thermodynamics, phase transitions, and a novel C-metric solution, were shown in \cite{Kubiznak:2023emm}.

The aim of the present paper is to study 
optical properties of RegMax black holes, such as photon spheres, gravitational red-shift, or the black hole shadow. Such an investigation is especially interesting because of the birefringence phenomena present in these spacetimes. As we shall see, one of the electromagnetic modes propagates along the background geometry, while the other mode follows the effective metric. In consequence, such black holes for example admit two photon spheres. 
Apart from astrophysical applications, we shall also use the opportunity to investigate further a potential connection between the black hole thermodynamic phase transitions and the {\em Lyapunov exponents} for unstable circular geodesics, as suggested in recent works  
\cite{Zhang:2019tzi, Guo:2022kio, Yang:2023hci, Lyu:2023sih, Kumara:2024obd}
(see also \cite{Cai:2021uov} for an alternative  investigation). 
This might be especially interesting for RegMax AdS black holes, which are known to  have involved thermodynamic behaviour in both the canonical and grandcanonical ensembles, including various phase transitions and critical points \cite{Kubiznak:2023emm}.\footnote{ 
Although Lyapunov exponents are primarily used to analyze dynamics of chaotic systems and to determine divergence/convergence of nearby trajectories in the phase space, multiple other uses were presented in the literature, see, e.g. \cite{Cardoso:2008bp}  for a study of quasinormal modes, or \cite{Hadar:2022xag, Kapec:2022dvc} for a novel holographic conjecture.} 
Such Lyapunov exponents are potentially measurable by future space-based detectors, e.g. \cite{Johnson:2019ljv, Gralla:2020srx}.

Our paper is organized as follows. In the next section, we review the 
basic properties of NLE theories and the associated birefringence, and introduce the  RegMax theory. The corresponding RegMax AdS black hole solutions, together with their basic characteristics, are summarized in Sec.~\ref{sec:bhsolution}. 
Section \ref{sec:geo} is devoted to the study of geodesics and Lyapunov exponents. Optical properties of asymptotically flat RegMax black holes are studied in Sec.~\ref{sec:optics}.  The connection between thermodynamic 
phase transitions of AdS RegMax black holes and the corresponding Lyapunov exponents in both canonical and grand-canonical ensembles is investigated in Sec.~\ref{sec:TDs}.  
We conclude in Sec.~\ref{sec:theend}. App.~\ref{A} summarizes a derivation of the central formula for Lyapunov exponents used in the main text.

\section{Introducing theories of NLE}\label{sec:one}

\subsection{RegMax theory}
Theories of NLE generalize Maxwell's theory by allowing their Lagrangian to be an `arbitrary' function of the two electromagnetic invariants:
\be 
\mathcal{S} = \frac{1}{2} F_{\mu\nu} F^{\mu\nu}\,,\quad 
\mathcal{P} = \frac{1}{2} F_{\mu\nu} \left(*F^{\mu\nu}\right)\,,
\ee
constructed from the corresponding electromagnetic field strength $F_{\mu\nu}$, or in terms of the vector potential $A_\mu$,
$F_{\mu\nu}=\partial_\mu A_\nu-\partial_\nu A_\mu$.
RegMax belongs to the {\em restricted class} of NLEs where only the invariant $\mathcal{S}$ is taken into account, i.e. $\mathcal{L}=\mathcal{L}(\mathcal{S})$.
Namely, its Lagrangian reads \cite{newLagrangian, Kubiznak:2023emm}:
\ba
\mathcal{L}&=&-2\alpha^4\left(1-3\ln(1-s)+\frac{s^3+3s^2-4s-2}{2(1-s)}\right)\,,\nonumber\\
s&=&{\sqrt[4]{-\frac{\mathcal{S}}{\alpha^4}}}\,.
\ea
The theory is characterized by a dimension-full parameter $\alpha$, whose square has a dimension of inverse length, and the  Maxwell theory is recovered upon the limit 
\be 
\alpha\to \infty\,. 
\ee
Upon denoting 
\begin{eqnarray}
D_{\mu\nu} \equiv \frac{\partial{\mathcal{L}}}{\partial{F^{\mu\nu}}}=2{\cal L}_{\cal S} F_{\mu\nu}\,,
\end{eqnarray}
where ${\cal L}_{\cal S}\equiv\frac{\partial{\mathcal{L}}}{\partial{\mathcal{S}}}, {\cal L}_{\cal S S}\equiv \frac{\partial^2\!{\mathcal{L}}}{\partial{\mathcal{S}}\partial{\mathcal{S}}}$ and so on, 
the generalized (vacuum) Maxwell equations read
\begin{eqnarray}
\dd F=0\,,~~   \dd *D=0\,. \label{eq:mxwll}
\end{eqnarray}
When minimally coupled to gravity, the corresponding Einstein equations, endowed with a (negative) cosmological constant $\Lambda=-3/\ell^2$, read
\be 
G_{\mu\nu}+\Lambda g_{\mu\nu}=8\pi T_{\mu\nu}\,,
\ee 
setting the Newton's constant $G=1$. Here, the (restricted) NLE energy momentum tensor $T_{\mu\nu}$ takes the following form:
\begin{eqnarray}
T^{\mu\nu}=-\frac{1}{4\pi}\bigl(2{\cal L}_{\cal S}F^{\mu\sigma}{F^{\nu}}_{\sigma}-\mathcal{L}g^{\mu\nu}\bigr)\,.
\end{eqnarray}

\subsection{Birefringence}

A characteristic property of all NLE theories, apart from the Maxwell and Born--Infeld cases, is the presence of {\em birefringence}. Namely, in the geometric optics approximation the two modes, corresponding to the two degrees of freedom encoded in the field, effectively propagate with respect to two distinct metrics. These may depend on the actual NLE field distribution, and can be found by describing the characteristic surfaces of propagation of field discontinuities  \cite{Novello:1999pg}.

Namely, let $\Sigma=\mbox{const}.$ be a wavefront surface and 
\be
k_\mu=\partial_\mu \Sigma
\ee
be the corresponding wave 1-form. Then, for a given mode one can construct an effective geometry $g_{\mbox{\tiny eff}}^{\mu\nu}$, with respect to which $k_\mu$ becomes null:
\be
g_{\mbox{\tiny eff}}^{\mu\nu} k_\mu k_\nu=0\,.
\ee
Defining further the inverse effective geometry $g^{\mbox{\tiny eff}}_{\mu\nu}$ by the standard relation
\be
g_{\mbox{\tiny eff}}^{\alpha\gamma}
g^{\mbox{\tiny eff}}_{\gamma\beta}=\delta^\alpha_\beta\,,
\ee
one finds \cite{Novello:1999pg} that the photon path, $x^\mu(\lambda)$,  
follows null geodesics w.r.t. the effective metric, namely, 
\be
\delta \int ds_{\mbox{\tiny eff}}=0\,.
\ee
In particular, defining 
\be
k^\mu\equiv g_{\mbox{\tiny eff}}^{\mu\nu} k_\nu=\frac{dx^\mu}{d\lambda}\,,
\ee
(with a proper choice of the affine parameter $\lambda$), it obeys the null geodesic equation
\be
k^{\alpha} \nabla_\alpha^{\mbox{\tiny eff}} k^\mu=0\,,
\ee
where $\nabla^{\mbox{\tiny eff}}$ is compatible with the effective metric.

It turns out that in the special case of the restricted class of NLE theories, $\mathcal{L}=\mathcal{L}(\mathcal{S})$, one mode propagates with respect to the standard {\em background metric} $g^{\mu\nu}$, while the other one follows the {\em effective metric} \cite{Novello:1999pg}:
\begin{eqnarray}\label{eff}
g_{\mbox{\tiny eff}}^{\mu\nu}=g^{\mu\nu}+\frac{2{\cal L}_{\cal SS}}{{\cal L}_{\cal S}}F^{\mu\alpha}{F^{\nu}}_{\alpha}\,. \label{eq:geff}
\end{eqnarray}
Note that the effective metric is defined up to a conformal rescaling -- all conformally related metrics yield the same null geodesics,  e.g. 
\cite{Wald:1984rg}.
In general, the birefringence in NLEs may lead to various pathologies, such as the existence of optical horizons, or closed lightlike curves, e.g. \cite{Novello:1999pg}.

\section{Charged black hole solution}\label{sec:bhsolution}
\subsection{Solution}
The charged black hole solution in the RegMax theory has been constructed in \cite{Kubiznak:2023emm}. It takes the following standard form: 
\begin{eqnarray}
g=g_{\mu\nu}\dd x^\mu \dd x^\nu=-f\dd t^2 +\frac{\dd r^2}{f}+r^{2}\dd \Omega^2\,,  \label{metric}
\end{eqnarray}
where $\dd \Omega^2=\dd \theta^2+\sin^2\!\theta \dd \varphi^2$, and the metric function $f$ reads 
\begin{align}
f=\,&1-2\alpha^{2}|Q|+\frac{4\alpha |Q|^{3/2}-6m}{3r}+\frac{r^2}{\ell^2}\nonumber \\
&+4\alpha^{3}r\sqrt{|Q|}-4r^2\alpha^4\ln\left(1+\frac{\sqrt{|Q|}}{r\alpha}\right)\label{eq:f}\\
=\,&{1-\frac{2m}{r}+\frac{Q^2}{r^2}+\frac{r^2}{\ell^2}+\frac{Q^2}{r^2}\sum_{n=1}^{\infty}\frac{4}{n+4}\Bigl(\frac{-\sqrt{|Q|}}{\alpha r}\Bigr)^n \,.\nonumber}
\end{align}
The metric is accompanied by the following vector potential:
\begin{eqnarray}
A=-\frac{Q\alpha}{r\alpha+\sqrt{\abs{Q}}} \dd t\,, \label{eq:potential}
\end{eqnarray}
which yields the corresponding field strength
\begin{eqnarray}
F= \mathcal{E} dr \wedge \dd t\,,\quad
\mathcal{E} \equiv \frac{Q\alpha^2}{(r\alpha+\sqrt{\abs{Q}})^2}\,,  \label{eq:eli}
\end{eqnarray}
and is characterized by the following two invariants:
\begin{eqnarray}
\mathcal{S}=-\mathcal{E}^2\,,\quad \mathcal{P}=0\,.    
\end{eqnarray}

\begin{figure}
\begin{center}
    \includegraphics[scale=0.67]{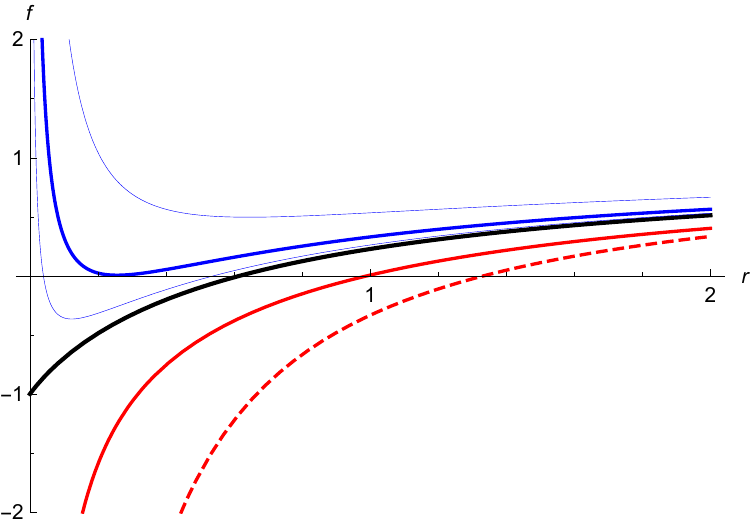}
	\caption{{\textbf{Black hole horizons.} We display the behavior of the metric function $f(r)$ for fixed values $\alpha=1,Q=1$ and $\Lambda=0$, as it depends on the mass parameter $m$. Namely, from top to bottom, we set $m=0.5137$, $m=0.6156$, $m=0.6497$, $m=m_{\text{marginal}}=2/3$ and $m=0.7771$; pure Schwarzschild with $m=2/3$ is shown in dashed red. We observe two distinct regimes:  the Schwarzschild-like regime denoted by red curves that occurs for $m>m_{\text{marginal}}$, and the Reissner-Nordström-like regime, for $m<m_{\text{marginal}}$, indicated by blue curves. The two regimes are separated by the marginal case, denoted by a solid black curve.
  }}
\label{fig:fm}
\end{center}
\end{figure}

The above solution is static, spherically symmetric, and singular at $r=0$.
In a certain range of parameters $\{m,Q,\alpha, \ell\}$, it describes a charged black hole, 
with the horizon radius $r_+$, given by the largest root of $f(r_+)=0$. 
In particular, switching off for a moment the cosmological constant, $\Lambda=0$, 
we display the mass dependence of $f=f(r)$ in Fig.~\ref{fig:fm} (cf. Fig.~2 in \cite{Kubiznak:2023emm} for the case of non-trivial $\Lambda$). 
For small masses $m$, the behavior of $f(r)$ resembles that of the Reissner--Nordstr{\"o}m solution. Namely, as the mass increases from zero, we move from having no roots corresponding to a naked singularity, to one degenerate root of the extremal black hole, and finally to two roots characteristic of a non-extremal black hole with inner and outer horizons. For even larger masses, 
\begin{align}
m>m_{\mbox{\tiny marginal}}=\frac{2\alpha\abs{Q}^{3/2}}{3}\,
\end{align}
the behaviour of the metric function switches from the Reissner--Nordstr{\"o}m mode to the Schwarzschild-like mode, characterized by the existence of a single non-extremal horizon, see Fig.~~\ref{fig:fm}.

\subsection{Physical properties}
As demonstrated in \cite{Kubiznak:2023emm}, the above solution is characterized by 
the following physical properties: 
the asymptotic mass $M$, and the electric charge $Q$:
\be
M=m\,,\quad Q=\frac{1}{4\pi}\int_{S^2}*D\,,   \label{Q RegMax}
\ee
the black hole temperature $T$, and entropy $S$:
\ba
T&=&\frac{f'(r_+)}{4\pi}\nonumber\\
&=&\frac{\alpha r_+(6|Q|\alpha^2+1)-2|Q|^{3/2}\alpha^2+\sqrt{|Q|}(1+12\alpha^4r_+^2)}{4\pi r_+(\alpha r_++\sqrt{|Q|})}\nonumber\\
&&-\frac{3r_+\alpha^4}{\pi}\log\bigl(1+\frac{\sqrt{|Q|}}{r_+\alpha}\bigr)+\frac{3r_+}{4\pi \ell^2}\,,\label{Tform}\\
S&=&\frac{\mbox{Area}}{4}=\pi r_+^2\,,
\ea
and the electrostatic potential:
\be\label{phiTD}
\phi=-\xi\cdot A\Bigr|_{r=r_+}=\frac{\alpha Q}{\alpha r_++\sqrt{|Q|}}\,, 
\ee
where in the last formula we used that the horizon is generated by the Killing vector field $\xi=\partial_t$.
Finally, since the solution is asymptotically AdS, we can consider the corresponding pressure-volume term \cite{Kastor:2009wy, Kubiznak:2016qmn}, 
\be\label{Vform}
P=-\frac{\Lambda}{8\pi}=\frac{3}{8\pi \ell^2}\,,\quad  V=\Bigl(\frac{\partial M}{\partial P}\Bigr)_{S,Q,
\alpha}=\frac{4}{3}\pi r_+^3\,,
\ee
and the ``{\em $\alpha$-polarization potential}'' \cite{Gunasekaran:2012dq}
\ba
\mu_\alpha&=&\Bigl(\frac{\partial M}{\partial \alpha}\Bigr)_{S,Q,P}\nonumber\\
&=&-\frac{2}{3}\frac{2|Q|^{3/2}\alpha r_+-Q^2-12\alpha^3r_+^3\sqrt{|Q|}-6|Q|\alpha^2r_+^2}{r_+\alpha+\sqrt{|Q|}}\nonumber\\
&&-8\alpha^3 r_+^3\log\Bigl(1+\frac{\sqrt{|Q|}}{r_+\alpha}\Bigr)\,, 
\ea
reflecting the fact that $\alpha$ is a dimension-full quantity.  

It is then easy to verify that the above quantities obey the extended first law and the corresponding Smarr relation  \cite{Kubiznak:2023emm}:
\ba
\delta M&=&T\delta S+\phi \delta Q+V\delta P+\mu_\alpha \delta \alpha\,,\\ 
M&=&2TS+\phi Q-2VP-\frac{1}{2}\mu_\alpha \alpha\,.
\ea
Moreover, the corresponding canonical (fixed charge) and grandcanonical (fixed potential) ensembles feature various critical points and phase transitions, see 
\cite{Kubiznak:2023emm}. We shall return to this feature in Sec.~\ref{sec:TDs} where we demonstrate that the corresponding critical points can be `discovered' by studying the Lyapunov exponents of the RegMax photon trajectories.

\section{Geodesics}\label{sec:geo}
\subsection{Effective metric}
In what follows we want to study various optical properties of the above black holes. To this purpose let us first turn to describing geodesics in these spacetimes. Namely, we shall be interested in null geodesics in the effective geometry \eqref{eff}, describing the propagation of one mode of non-linear RegMax photons, as well as the timelike and null geodesics in the background metric, describing the motion of massive test objects, and the propagation of the second mode of non-linear photons and of other massless particles.

To simultaneously treat the background metric $g_{\mu\nu}$ and the effective metric $g_{\mu\nu}^{\mbox{\tiny eff}}$ (defined as the inverse to $g_{\mbox{\tiny eff}}^{\mu\nu}$), let us use a shorthand for both possibilities, namely a new metric 
\be
q=q_{\mu\nu}\dd x^\mu \dd x^\nu=
\begin{cases}
    \text{background metric}\ g\\
    \text{effective metric}\ g_{\mbox{\tiny eff}}
\end{cases}\,,
\ee
dependent on which situation we want to study. That is, $q=g$, 
\eqref{metric} for the case of the background metric, and, inverting \eqref{eff},
\ba
q&=&g_{\mbox{\tiny eff}}=-fh \dd t^2+\frac{h}{f}\dd r^2+r^2 \dd\Omega^2\,,\nonumber\\
h&=&1-\frac{\sqrt{|{\cal E}|}}{\alpha}=
\frac{\alpha r}{\alpha r+\sqrt{|Q|}}\in (0,1)\,
\ea
for the case of the effective metric. As we approach spatial infinity, $h \to 1$, and both  metrics $q_{\mu\nu}$ are asymptotically flat/AdS.\footnote{In particular, this implies that we can use the same Killing vectors $\partial_t$ and $\partial_\varphi$ to denote the asymptotic symmetries for both these metrics.}

Because of the spherical symmetry, the motion of the massive and massless particles is effectively 3-dimensional, and without loss of generality takes place in the $\theta=\pi/2$ plane. The corresponding effective Lagrangian then reads:
\begin{align}
\mathcal{L}=\frac{1}{2}\left(q_{tt}\dot{t}^2 +q_{rr}\dot{r}^2+r^{2}\dot \varphi^2  \right)\,, \label{Lagrangian of test particles}
\end{align}
where the dot denotes a derivative w.r.t. an affine parameter characterizing the geodesic.
Employing further the static and axisymmetric Killing vectors
\be\label{KVs}
 \xi=\partial_t\,,\quad \zeta=\partial_\varphi\,,
\ee
together with the normalization of the 4-velocity, $u^\mu=(\dot t, \dot r, 0,\dot \varphi)$, we have the following 3 integrals of motion:
\ba\label{integrals}
E&=&-\xi\cdot u=q_{tt}\dot t\,,\nonumber\\ L&=&\zeta\cdot u=r^2\dot \varphi\,,\\
\kappa&=&u^2=q_{tt}{\dot t}^2+q_{rr}{\dot r}^2+r^2{\dot \varphi}^2\,,\nonumber
\ea
where $\kappa=-1,0$ for timelike, null geodesics. Plugging the first two expressions into the last one, we then obtain the radial equation 
\be
{\dot r}^2+V_r=0\,,\quad V_r=\frac{1}{q_{rr}}\Bigl(\frac{E^2}{q_{tt}}+\frac{L^2}{r^2}-\kappa\Bigr)\,,\label{Veff}
\ee
governing the motion of test particles in the equatorial plane of the spherically symmetric spacetime with metric $q_{\mu\nu}$.

\subsection{Circular Geodesics}

\begin{figure}
\begin{center}
	\includegraphics[scale=0.67]{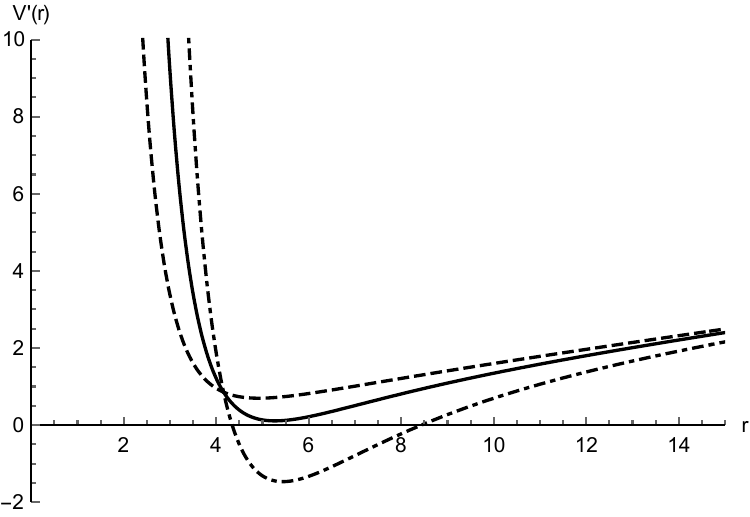}
	\caption{\textbf{Effective potential: timelike geodesics.} In this figure, we display the dependence of the first derivative of the effective potential on particle's angular momentum $L$; a $V_r'-L$ diagram. The dot-dashed line corresponds to $L=30$, the black line to $L=18$, and the dashed one to $L=10$, with other parameters taking values $\alpha=1$, $Q=1$, $P=0.01$, $r_+=2$. 
   As is clear from the plot, function $V_r'(r)$ admits non-trivial roots only for sufficiently large $L$; in that case, the larger root corresponds to the stable circular trajectory and the smaller root to the unstable circular trajectory.}
\label{fig:V-L}
\end{center}
\end{figure}

Let us first focus on circular geodesics. These are given by 
\begin{align}
    V_r(r_c)=&0\,,\label{circular geodesic condition 1}\\
    V_r'(r_c)=&0\,,\label{circular geodesic condition 2}
\end{align}
where $r_c$ is the corresponding radius. Stable/unstable orbits are characterized by $V_r''(r_c)$ being positive/negative, respectively. 

\begin{figure}
\begin{center}
	\includegraphics[scale=0.67]{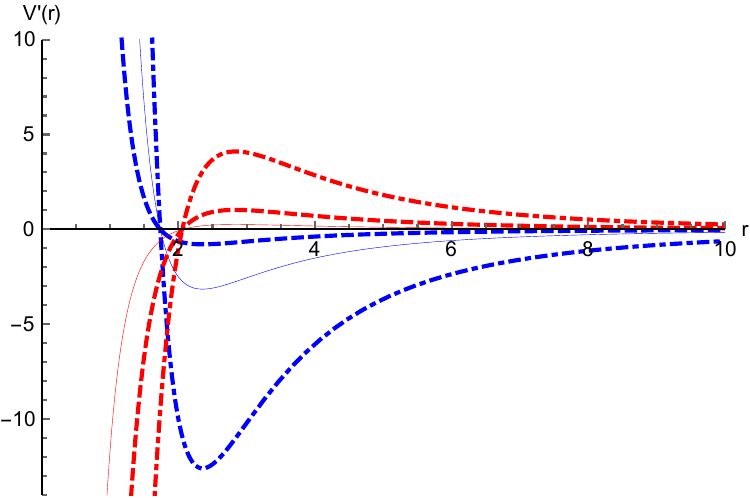}
	\caption{\textbf{Effective potential: null geodesics.} This figure depicts the dependence of the first derivative of the effective potential $V_r'(r)$ on angular momentum $L$ for null geodesics. Red curves correspond to the effective metric, blue ones to the background metric, values of angular momentum are $L=5$ for solid curves, $L=10$ for dashed curves and $L=20$ for dashed-dot curves; other parameters take values $\alpha=1,Q=1,r_+=1,P=0.01$. We can see that $L$ only affects prominence of local maxima/minima while the root of $V_r'(r)$ stays the same -- the circular radius $r_c$ is independent of $L$ in the case of null geodesics. Moreover, we observe that $r_c^{\mbox{\tiny eff}}>r_c^{\mbox{\tiny bcg}}$. }
\label{fig:V-L-null}
\end{center}
\end{figure}

Let us first consider timelike 
$(\kappa\neq 0$) geodesics.  The dependence of the first derivative of the effective potential $V_r'$ on particle's angular momentum $L$ is displayed in Fig.~\ref{fig:V-L}. We see that for sufficiently large $L$, there are two possible roots; with the smaller one corresponding to unstable circular orbits, and  the larger one giving the stable ones.  More concretely, the conditions \eqref{circular geodesic condition 1} and \eqref{circular geodesic condition 2} yield
\ba
    E^2&=&\frac{-2 g_{tt}^2}{2g_{tt}-r g_{tt}'}\Bigl|_{r=r_c}\,,\label{eq:Ecircular}\\
    L^2&=&\frac{r^3g_{tt}'}{2g_{tt}-r g_{tt}'}\Bigl|_{r=r_c}\,\label{eq:Lcircular}\,,
\ea    
or, introducing the impact parameter $b$,
\be 
    b\equiv\frac{L}{E}=\pm\sqrt{-\frac{r^3g_{tt}'}{2g_{tt}^2}}\Bigl|_{r=r_c}\,.\label{def impact parameter}
\ee

Turning next to the null geodesics $(\kappa=0$), the derivative of the effective potential is displayed in Fig.~\ref{fig:V-L-null}. As obvious, the  circular (photon sphere) radius is independent of $L$; it can be determined from 
\be
 q_{tt}'(r_c)=\frac{2q_{tt}(r_c)}{r_c}\label{null derivative reduction}\,,
\ee
and corresponds to (real) $b_\gamma$, given by 
\be
 b_{\gamma}\equiv \frac{L_\gamma}{E_\gamma}=\pm\sqrt{-\frac{r_c^2}{q_{tt}(r_c)}}\,.\label{def null impact parameter}
\ee
Here, $\pm$ corresponds to left, right moving photons, moving on the photon sphere. 
For a non-extremal RegMax black hole, Eq.~\eqref{null derivative reduction} yields 
exactly two {\em photon spheres} outside the black hole horizon. One is connected with the background metric (where photons of one polarization orbit), and the second one is due to the effective metric (where photons of the second polarization orbit).
It would be interesting to see whether this could give rise to some observable signatures -- 
see, e.g. \cite{chen2023interferometric} for a recent discussion of interferometric signatures of black holes with multiple photon spheres.\footnote{Considering generic beam of electromagnetic radiation, and neglecting its backreaction on the geometry and the background electromagnetic field, one can always split the beam into the corresponding two  polarizations, each propagating according to its own effective metric.

Note also that  
the horizon of an extremal black hole trivially satisfies the condition \eqref{null derivative reduction} for both electromagnetic modes. 
The extremal horizon therefore becomes another photon sphere that also corresponds to a minimum of the effective potential $V_r$, i.e. stable circular null geodesics. Such behavior seems generic for extremal black holes, e.g. \cite{Pradhan:2010ws, Khoo:2016xqv}.
For restricted NLE's, this photon sphere is `universal' -- present for both electromagnetic modes.  
}

While suffering from birefringence, the propagation of photons in RegMax theory is rather `nice and intuitive' -- no pathologies, such as the existence of electromagnetic trapped surfaces, or closed lightlike curves, that may arise  \cite{Novello:1999pg}  in more complicated theories of NLE,
occur for spherical black holes in RegMax theory.

In what follows, it would be useful to find algebraic expressions for the radii of the above circular geodesics.
Unfortunately, due to the 
logarithmic term in the metric function $f$, this becomes difficult for the case of timelike geodesics or null effective geodesics. (When needed, we will have to determine these radii numerically, using Eqs.~\eqref{circular geodesic condition 1} and \eqref{circular geodesic condition 2}.) The situation is different for null geodesics in the background metric, for which the logarithmic term in \eqref{null derivative reduction} vanishes, and its roots can be found explicitly:
\begin{align}
\rho_{\pm} =& \frac{3m\alpha-\sqrt{|Q|}}{2\alpha}\pm \frac{ \sqrt{|Q|-6m\sqrt{|Q|}+\alpha^2(9m^2-8Q^2)} }{2\alpha} \label{eq:rc}\nonumber\\
=&\frac{1}{2}\lr{(}{3m\pm\sqrt{9m^2-8Q^2}}{)}\nonumber\\&+\frac{\sqrt{|Q|}}{2\alpha}\lr{(}{-1\pm\frac{3m}{\sqrt{9m^2-8Q^2}}}{)}+O(\frac{1}{\alpha^2})\nonumber\\
=&(3\pm3)\frac{m}{2}+(-1\pm1)\frac{\sqrt{|Q|}}{2\alpha}\nonumber\\&-\lr{(}{\pm\frac{2Q^2}{3m}}{)}+\lr{(}{\pm\frac{2|Q|^{5/2}}{9\alpha m^2}}{)}+O(|Q|^3)\,.
\end{align}
The root $\rho_-$ has to be excluded, as it lies below the horizon and corresponds to imaginary $b_\gamma$. 
It is the  $\rho_+$ root that corresponds to the photon sphere radius outside the horizon:  
\be\label{rc_formula}
   r_c\equiv\rho_+\,.
\ee
The first term in the expansion for large $\alpha$ 
agrees with the Reissner-Nordstr\"om photon sphere radius, while the Schwarzschild photon sphere radius is recovered upon setting $Q\to 0$ in the latter expansion.

\subsection{Lyapunov Exponents}
Test particles near the black hole can move along different trajectories. Lyapunov exponents can measure how divergent or convergent these trajectories are in the phase space. Namely, positive value of Lyapunov exponent indicates that nearby trajectories are divergent and thus depend strongly on initial conditions. Motion of particles then bears a close resemblance to chaotic systems \cite{Cardoso:2008bp}.
However, Lyapunov exponents proved to be useful beyond the realm of chaotic systems investigation. In \cite{Cardoso:2008bp}, they were used for studying quasinormal modes. It was shown that in asymptotically flat, spherically symmetric black hole spacetimes, quasinormal modes in geometrical optics approximation can be understood as slowly leaking particles trapped at the unstable circular null geodesic with leaking timescale given by the principal Lyapunov exponent.
{More recently,  
a proposal for a holography of a photon sphere was made in \cite{Hadar:2022xag, Kapec:2022dvc}. Namely, it was suggested that the  classical Lyapunov exponents are dual to the quantum Ruelle resonances describing the late-time approach to thermal equilibrium of the quantum microstate dual to a given asymptotically flat black hole. Since such Lyapunov exponents are potentially measurable by the near future space-based detectors,  e.g. \cite{Johnson:2019ljv, Gralla:2020srx}, this could be a  starting point for a bottom-up approach to holography for astrophysical black holes.
}

There were also recent attempts
at using the geodesic Lyapunov exponents
to analyse thermodynamic phase transitions in black hole spacetimes, e.g. 
\cite{Zhang:2019tzi, Guo:2022kio, Yang:2023hci, Lyu:2023sih, Kumara:2024obd}. In our paper, we exploit Lyapunov exponents for this purpose. Concentrating on trajectories that are nearby to an unstable circular trajectory at radius $r_c$, the Lyapunov exponent takes the following form (see App.~\ref{A} for a derivation of this formula):
\begin{eqnarray}
\lambda=\sqrt{-\frac{V_{r}^{''}}{2\dot{t}^2}}\biggl|_{r=r_c}\,. \label{def lambda}
\end{eqnarray}
To evaluate this for timelike geodesics, we employ the effective potential \eqref{Veff}, together with \eqref{integrals}, and eliminate $E^2$ and $L^2$ by using \eqref{eq:Ecircular} and \eqref{eq:Lcircular}, giving:
\be\label{Lambda_timelike}
\lambda=\sqrt{\frac{g_{tt}''}{2g_{rr}}-\frac{(g_{tt}')^2}{g_{tt} g_{rr}}+\frac{3g_{tt}'}{2rg_{rr}}}{\Bigl|}_{r=r_c}\,.
\ee
Similarly, for null geodesics, we employ  \eqref{Veff}, \eqref{integrals}, and  \eqref{null derivative reduction} together with \eqref{def null impact parameter}, to obtain
\be
\lambda_\gamma=\sqrt{\frac{q_{tt}''}{2q_{rr}}-\frac{q_{tt}}{r^2q_{rr}}}\Bigl|_{r=r_c}\,.
\ee
In particular, using the explicit form of our metrics, the photon Lyapunov exponents take the following explicit form: 
\be
\lambda_1=\sqrt{\frac{f^2}{r^2}-\frac{f(fh)''}{2h}}\biggl|_{r=r_c}\,,\label{lambda1}
\ee
for the effective metric, and 
\be
\lambda_2=\sqrt{\frac{f^2}{r^2}-\frac{ff''}{2}}\biggl|_{r=r_c}\,,\label{lambda2}
\ee
for the background metric.

\section{Optical phenomena}\label{sec:optics}
Let us now study some optical phenomena associated with RegMax black holes. These are particularly interesting due to birefringence in our theory. In this section we focus on astrophysically more relevant asymptotically flat black holes, setting $\Lambda=0$.

\subsection{Phase velocities}
\begin{figure}
\begin{center}
	\includegraphics[scale=0.67]{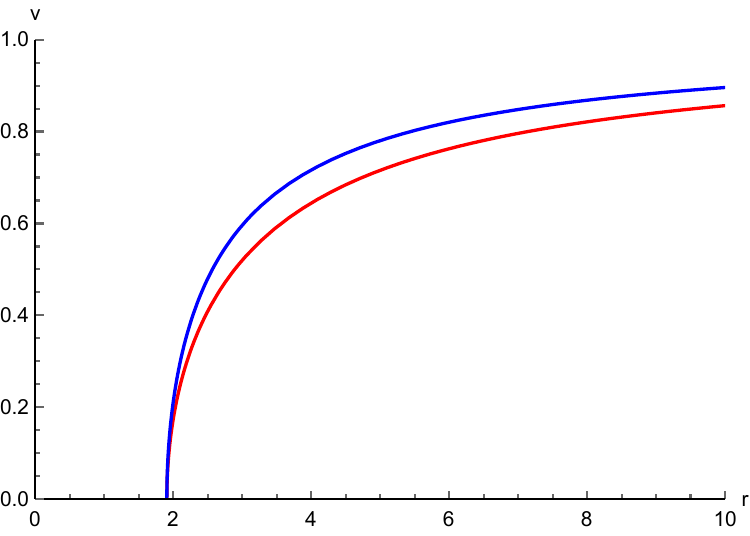}
	\caption{\textbf{Phase velocities:} In this figure, we can clearly see the birefringence for the case of motion in the azimuthal direction; red curve stands for $v^{(\varphi)}_{\mbox{\tiny eff}}$, blue curve for $v^{(\varphi)}$. In the case of motion in radial direction, phase velocities for both effective and background metric are the same and coincide with $v^{(\varphi)}$. Here, we have set $m=1, Q=0.5, \alpha=0.75$. Note that the curves emerge from $r=1.90567$, which is the horizon radius.}
\label{fig:phasevel}
\end{center}
\end{figure}

Let's 
consider the light wave vector  
\be
k_\mu=(-\omega, k_i)\,,\quad k_i=(k_r,k_\theta, k_\varphi)\,,
\ee
with $\omega>0$, and define the {\em phase velocity} as \cite{Liberati:2000mp} (see also 
\cite{Liberati:2001sd, Shore:2002gn}):
\be
v\equiv \frac{\omega}{|\Vec{k}|},
\ee
where we defined the spatial magnitude $|\Vec k|$ in terms of the background metric, that is $|\Vec{k}|=\sqrt{g^{ij}k_ik_j}$. Such $v$ carries the meaning of velocity of the surface of constant phase in a particular direction $\Vec{k}$. Since we have 
\be 
0=q^{\mu\nu}k_\mu k_\nu=\omega^2 q^{tt}+q^{ij}k_i k_j\,,
\ee 
we can easily evaluate the phase velocity for purely radial or purely angular wave vectors. Namely, 
for radial geodesics we have $k_\mu=(-\omega, k_r,0,0)$. Then $|\Vec k|^2=\omega^2/f$ for both effective and background metrics, and we find 
\be
v^{(r)}=\pm \sqrt{f}\,.
\ee
for both types of modes.
Obviously, for purely radial motion, there is no birefringence effect on phase velocities. The situation is, however, different when angular motion comes into play; if the wave vector has non-zero $k_\theta$ or $k_\varphi$ component, phase velocities will differ for the two metrics. In particular, consider the motion in the $\varphi$-direction, that is, 
$k_\mu=(-\omega, 0, 0, k_\varphi)$.
Then $|\Vec k|^2=\omega^2/(fh)$ for the `effective mode' and $|\Vec k|^2=\omega^2/f$ for the `background mode'. Thus, in this case, we get two different phase velocities, namely 
\be\label{eq: phase velocity phi eff}
v^{(\varphi)}_{\mbox{\tiny eff}}=\pm \sqrt{fh}\,,
\ee
for the effective mode, and  
\be\label{eq: phase velocity phi bcg}
v^{(\varphi)}=\pm \sqrt{f}\,.
\ee
for the background mode, and similarly for the motion in the $\theta$-direction. Note that in this case, and since  $h\in\left(0,1\right)$, 
the angular motion wave-fronts in the effective metric travel slower than the modes in the background metric. We display the corresponding phase velocities as a function of radial coordinate $r$ in Fig.~\ref{fig:phasevel}. 

To summarize, birefringence is observable every time the wave vector has a non-zero angular component $k_\theta$ or $k_\varphi$. On the other hand, if angular components vanish and we are left only with the radial one, phase velocities for both metrics are the same and there is no splitting. This is a feature of any restricted theory of NLE with ${\cal L}={\cal L}({\cal S})$. In particular, this will be true on the horizon -- radial photons emitted from the horizon propagate in the same way as gravitons.

Recently, there was some speculation,  e.g. \cite{Hajian:2020dcq}, that 
if such velocities were different, it might lead to a modification of the Hawking temperature. However, this is clearly not the case for 
black holes in non-linear electrodynamics, for which the  first law should then take on the standard form, with Hawking temperature identified with surface gravity. In particular, this will be true for regular black holes identified as solutions corresponding to a magnetic monopole in due non-linear electrodynamics.

\subsection{Light trajectories}
We have seen, that due to  birefringence, there are two photon spheres (at two different radii) around RegMax black holes. When slightly perturbed inwards, such photons will inspiral into the black hole. Here we describe their trajectories in the equatorial plane. \\

The radial part of the inspiral is described by the effective potential \eqref{Veff} (with $\kappa=0$), while the azimuthal part is given by the second equation \eqref{integrals}.  
For the trajectory of the inspiral, this yields 
\ba \label{drdphi}
\frac{dr}{d\varphi}&=&\frac{\dot r}{\dot \varphi}=-r^2\sqrt{-\frac{V_r}{L^2}}\nonumber\\
&=&-r^2\sqrt{-\frac{1}{q_{rr}}\Bigl(\frac{1}{r^2}-\frac{1}{b_\gamma^2 q_{tt}}\Bigr)}\,,
\ea 
where, for the impact parameter $b_\gamma$ we plug the unstable circular orbit value, \eqref{def null impact parameter}. Here, $r_c$ is given by \eqref{rc_formula} for the case of photons following the background geometry, and $r_c$ has to be determined numerically for the photons following the effective metric. 

\begin{figure*}
\subfloat[\label{fig:PTfixRp}]{
	\includegraphics[width=0.35\linewidth]{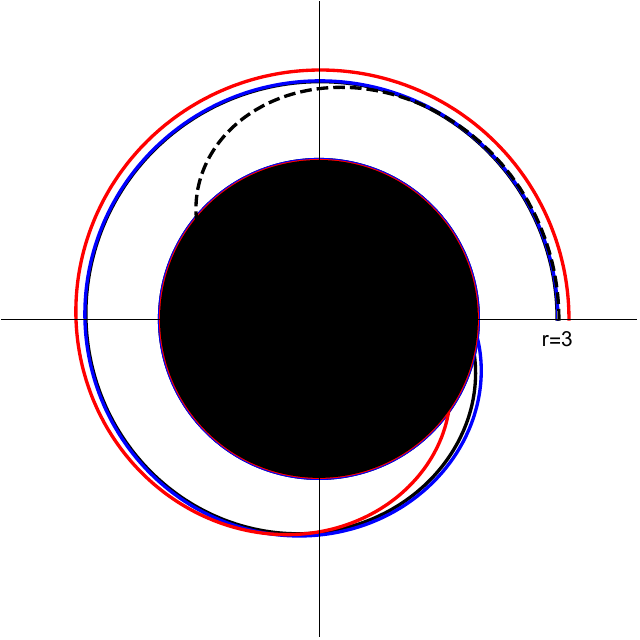}
}
\subfloat[\label{fig:PTfixRc}]{
	\includegraphics[width=0.35\linewidth]{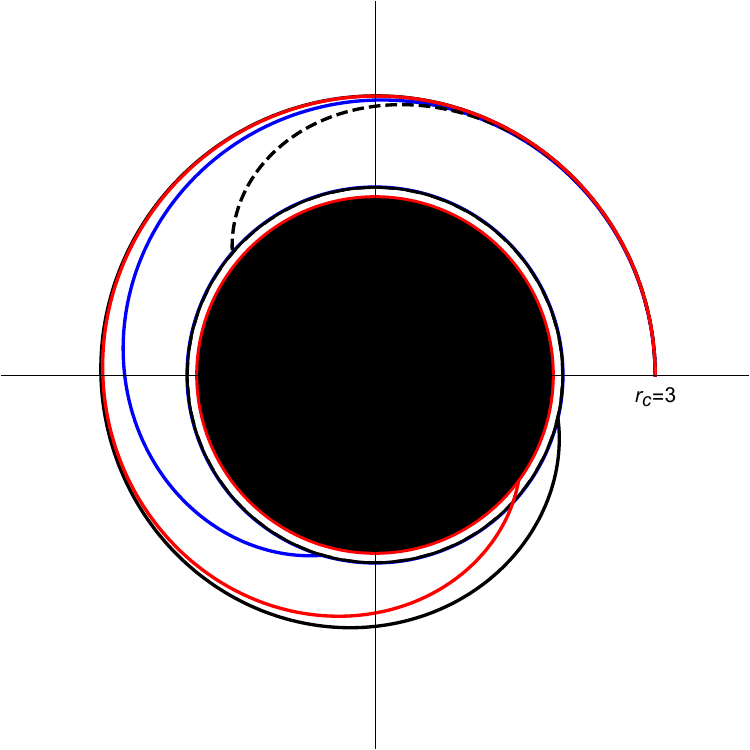}
}\\
\subfloat[\label{fig:PTfixISCO}]{
	\includegraphics[width=0.35\linewidth]{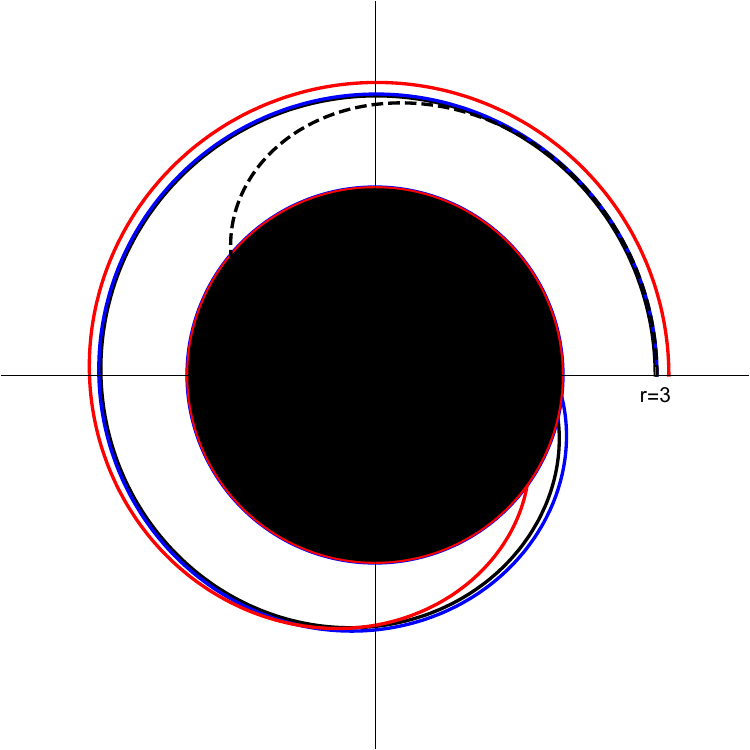}
}
\subfloat[\label{fig:PTfixM}]{
	\includegraphics[width=0.35\linewidth]{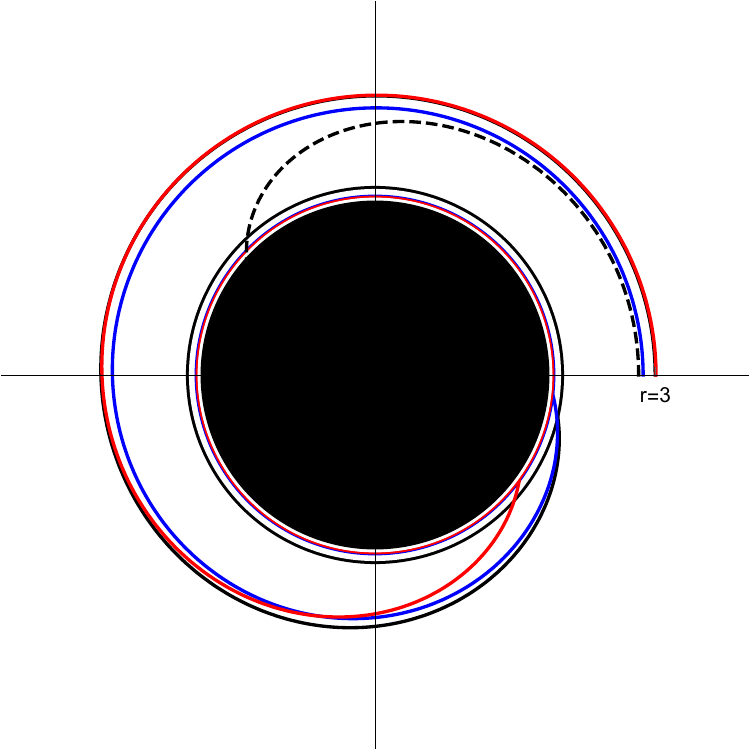}
}
\caption[]{\textbf{Light trajectories around black holes.} In this figure we display trajectories of light rays emerging from the vicinity of the photon sphere, as they inspiral towards the black hole for Schwarzschild (solid black), Reissner--Nordstr\"om (dashed black), and RegMax black holes in the case of effective (red) and background (blue) metrics. In all figures, the RegMax black holes are characterized by  $\alpha=1/\sqrt{2}$ and $Q=1/2$, the Reissner--Nordstr\"om black holes by $Q=1/2$, and the light ray is sent along the null geodesics just below the particular photon sphere, at $r_0=0.999 r_c$. Moreover, if different, we display the black hole horizons with circles of the corresponding colour. {\bf (a)} {\em Equal horizon radius.} In this subfigure we fix $r_+=2$ for all black holes. Other parameters are as follows: Schwarzschild ($r_c=3$, $M=1$), Reissner--Nordstr\"om ($r_c=3.0022, M=1.0625$), and RegMax ($M=1.0448$) with background ($r_c=3.010$) and effective ($r_c=3.1492$). {\bf (b)}  {\em Equal photon sphere radius}. In this subfigure all cases  share the same photon sphere radius  $r_c=3$; other parameters are: Schwarzschild ($r_+=2$, $M=1$), Reissner--Nordstr\"om ($r_+=1.997, M=1.055$), and RegMax background ($r_+=2.006$, $M=1.0417$) and effective ($r_+=1.9011$, $M=0.9970$). {\bf (c)}  {\em Equal ISCO radius} $r_{\text{\tiny ISCO}}=6$ for: Schwarzschild ($r_+=2$, $r_c=3$, $M=1$), Reissner--Nordstr\"om ($r_+=1.9977$, $r_c=3.0186$, $M=1.0614$), and RegMax background ($r_+=2.0063$, $r_c=3.0191$, $M=1.0478$) and effective ($r_+=1.9981$, $r_c=3.1463$, $M=1.0439$). {\bf (d)} {\em Equal mass} $M=1$ for: Schwarzschild ($r_+=2$, $r_c=3$), Reissner-Nordstr\"om ($r_+=1.8660$, $r_c=2.8229$), and RegMax ($r_+=1.9073$) for background ($r_c=2.8708$) and effective 
($r_c=3.0094$). 
    }
    \label{fig:photon trajectories}
\end{figure*}

We display the corresponding numerically constructed inspirals  
in the vicinity of RegMax black holes in Fig.~\ref{fig:photon trajectories},
where they are also compared with similar inspirals in Schwarzschild and 
Reissner--Nordström spacetimes. In these figures, in order to compare different black hole spacetimes, we choose to fix one black hole parameter, common to all spacetimes considered. 
Namely, we present four pictures of the inspirals -- choosing the same horizon radius for all black holes (Fig.~\ref{fig:photon trajectories}\subref{fig:PTfixRp}), the same photon sphere radius (Fig.~\ref{fig:photon trajectories}\subref{fig:PTfixRc}), black holes with the same ISCO radius (Fig.~\ref{fig:photon trajectories}\subref{fig:PTfixISCO}), and black holes with the same mass parameter 
(Fig.~\ref{fig:photon trajectories}\subref{fig:PTfixM}).

In particular, in Fig.~\ref{fig:photon trajectories}\subref{fig:PTfixRp}, we compare the inspirals for a fixed RegMax black hole for the effective (red) and background (blue) cases. We observe that, while the effective metric photon sphere is `bigger' (`lies outside' the background metric photon sphere), giving the effective photons `more space' to inspiral further,  the effective pull is `stronger' and so, in fact, the photons following the effective metric reach the horizon `sooner'. On the other hand, from Fig.~\ref{fig:photon trajectories}\subref{fig:PTfixRc}, where the inspiral is studied for the same fixed photon sphere radius, 
the background photons seem to inspiral `faster' than the effective ones. While this seems strange, there is no contradiction between these two figures, as in the later case, the black hole horizons are different (we are comparing two distinct RegMax black holes), with the `background' RegMax black hole bigger and more massive than the `effective' RegMax black hole.

When interpreting all these pictures, however, one needs to be a bit cautious. The curves in Fig. \ref{fig:photon trajectories} correspond to several \emph{different} spacetimes and/or metrics, and even in the cases where we take RegMax black holes with the same mass parameter and horizon radius, the meaning of the radial coordinate $r$ is not exactly the same in the effective and background cases. However, we have fixed the Weyl conformal factor so that all metrics governing the light behavior are written in the `area gauge', for which the radial coordinate is constructed in such a way that a sphere of radius $r$ has a surface given by $4\pi r^2$. In particular, this implies that a photon sphere at larger $r$ is `bigger' in the sense that it has larger area. This does not, though, imply that, for example, the proper distance to the horizon has to also be larger. 
Another ground where one must tread with care is the choice of the inward perturbation of $r_c$ in order to get the light inspiraling onto the black hole. In Fig.~\ref{fig:photon trajectories}, we always send the light signal from the same percentage of (respective) $r_c$ (namely at $r_0=0.999 r_c$). However, since the photon sphere radii are distinct for distinct metrics, the impact of this perturbation can also differ and some of the photons may experience stronger `kick' than others.

\begin{figure}
\begin{center}
	\includegraphics[scale=0.67]{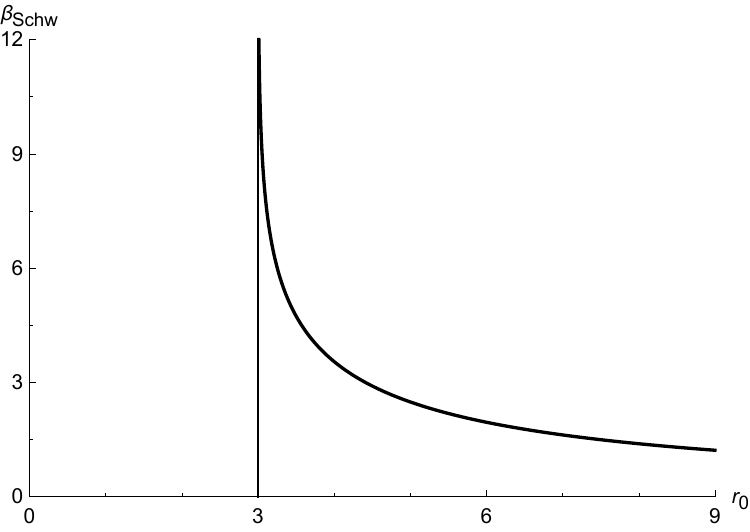}
	\caption{\textbf{Deflection angle for Schwarzschild.} The deflection angle $\beta(r_0)$ is calculated using formula \eqref{eq: deflection angle} for the Schwarzschild solution with $M=1$, i.e.,  $r_c=3$.
 }
\label{fig: deflection angle Schw}
\end{center}
\end{figure}

\begin{figure}
\begin{center}
	\includegraphics[scale=0.67]{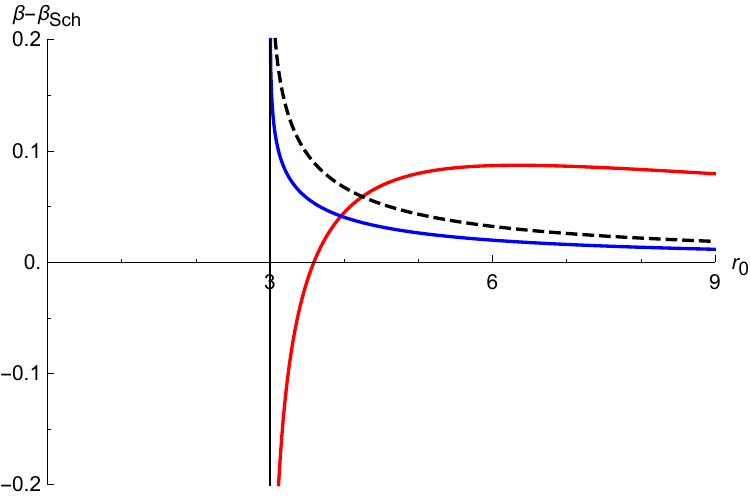}
	\caption{\textbf{Deflection angle for RegMax black hole.} The difference $\beta-\beta_{Sch}$, where $\beta_{Sch}$ is for reference plotted in Fig.~\ref{fig: deflection angle Schw}, is displayed for  background metric (blue), effective metric (red), and 
 Reissner--Nordstr\"om (dashed black) solutions with parameters fixed so that $r_c=3$, $Q=1/2$ $\alpha=1/\sqrt{2}$. Other parameters for the situation where $r_c=3$ are presented in Fig.~\ref{fig:PTfixRc}.
 }
\label{fig: deflection angle fix M}
\end{center}
\end{figure}

Having the above in mind, 
let us also calculate the deflection angle for the light passing our black hole at the closest approach at $r_0$. This is simply given by the following integral:
\begin{align}\label{eq: deflection angle}
\beta(r_0)=2\int_{r_0}^{\infty}\dv{\varphi}{r}\,\dd r\,,
\end{align}
where $\frac{d\varphi}{dr}$ is simply given by $1/\dv{r}{\varphi}$ from \eqref{drdphi} as the derivative of the inverse function. Since at the closest approach $\dot{r}=0$, so by the normalization of the null wave-four-vector $k^\mu$ \eqref{integrals}, $b_\gamma$ in \eqref{drdphi} is given by \eqref{impact parameter} with $r_e=r_0$. For reference, we display the deflection angle for Schwarzschild with $r_c=3$ in Fig.~\ref{fig: deflection angle Schw}. This is then compared to the RegMax deflection angles in the effective and background metrics in Fig.~\ref{fig: deflection angle fix M}.

\subsection{Gravitational redshift}
The gravitational redshift
determines the decrease of frequency of photons as measured 
by  
an `emitter' $e$ moving with four-velocity $U_e^\mu$, and 
an 
`observer' 
$o$ moving with $U_o^\mu$. These frequencies $\omega_i$, where $i\in \{e,o\}$, are determined by the projections of the photons' wave-four-vectors $k^{\mu}$ onto the respective four-velocities:
\begin{align}\label{eq: frequency}
    \omega_i=-\left(k_\mu U^\mu\right)_i.
\end{align}
The redshift is then defined as:
\be 
  z=\frac{\omega_e-\omega_o}{\omega_o}=\frac{\omega_e}{\omega_o}-1\,.
\ee
In what follows, we focus on determining
the redshift of a photon travelling along null geodesics in the equatorial plane emitted and observed on circular timelike geodesics orbiting the black hole, i.e., $U_i^\mu=(U_i^t,0,0,U_i^\varphi)$ and $k_\mu=(k_t,k_r,0,k_\varphi)$, in which case we can write:
\begin{align}
    1+z=\frac{\left(k_t U^t+k_\varphi U^\varphi\right)|_e}{\left(k_t U^t+k_\varphi U^\varphi\right)|_o} \,.\label{redshift1}
\end{align}
Employing further the two Killing vectors \eqref{KVs}, we have 
\begin{align}
    E_\gamma=&-q_{\mu\nu}\xi^\mu k^\nu=-k_t,\label{Egamma}\\
    L_\gamma=&q_{\mu\nu}\zeta^\mu k^\nu=q_{\varphi\varphi}k^{\varphi}=k_\varphi\,.\label{Lgamma}
\end{align}
Introducing the test particle angular velocities $\Omega_i=\frac{U_i^\varphi}{U_i^t}$ and again the impact parameter $b_\gamma=L_\gamma/E_\gamma$, and using that all three of these are integrals of motion, we can rewrite  \eqref{redshift1} as 
\be 
    1+z=\frac{\lr{(}{-U^t+b_\gamma U^\varphi}{)}|_e}{\lr{(}{-U^t+ b_\gamma U^\varphi}{)}|_o}
   = \frac{U_e^t}{U_o^t}\lr{(}{\frac{-1+ b_\gamma\Omega_e}{-1+ b_\gamma\Omega_o}}{)}\,.\label{eq: redshift not yet}
\ee
In particular, when a photon is emitted tangentially to the circular geodesic of the emitting test particle, we have $k_e^r=0$, from which, by the normalization of four-velocity of the null wave-four-vector $k^\mu$, \eqref{integrals} follows an additional relationship between $E^2$ and $L^2$:
\begin{align}
    \frac{L^2}{E^2}=-\frac{r_e^2}{q_{tt}(r_e)}\equiv b_\gamma^2\,.\label{impact parameter}
\end{align}
This effectively generalizes the formula \eqref{def null impact parameter} to this situation.

\begin{figure}
\begin{center}
	\includegraphics[width=\linewidth]{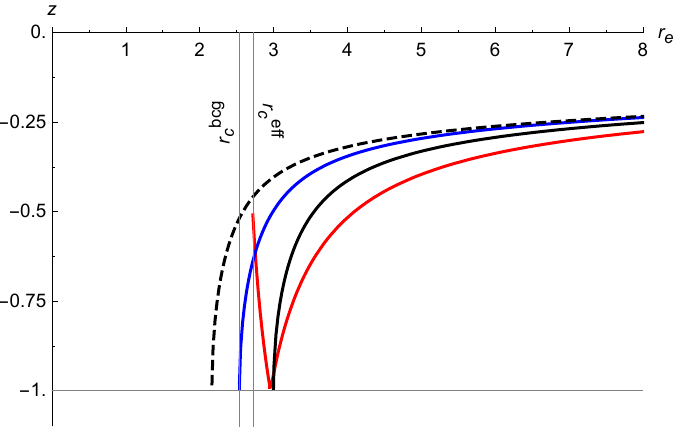}
	\caption{\textbf{Gravitational redshift: prograde emission.} In this picture, we display gravitational redshifts of photons emitted tangentially and in the direction of motion of a massive test particle moving on a circular orbit and observed by a static observer at infinity, as they depend  on emitter's radius $r_e$. The red curve corresponds to the RegMax ($M=1, Q=0.95, \alpha=1/\sqrt{2}$) effective metric, the blue curve to the RegMax background metric, solid black to Schwarzschild $(M=1)$, and dashed black to the Reissner--Nordstr\"om ($M=1, Q=0.95)$. 
 Evidently, the kinematic blueshift, due to the relativistic Doppler effect of the light being emitted in the direction of the orbiting test particle, `outcompetes' the effect of pure gravitational redshift. The point in the red line where $z=-1$ corresponds to the radius $r_s\approx2.9597$ where the photon following the effective metric `appears static' to the emitter and $\omega_e$ changes sign. The blue curve terminates at the background photon sphere radius $r_c^{\mbox{\tiny bcg}}$, while the red curve terminates at the effective photon sphere radius, $r_c^{\mbox{\tiny eff}}$.}
\label{fig:gredshift2}
\end{center}
\end{figure}

\begin{figure}
\begin{center}
	\includegraphics[scale=0.67]{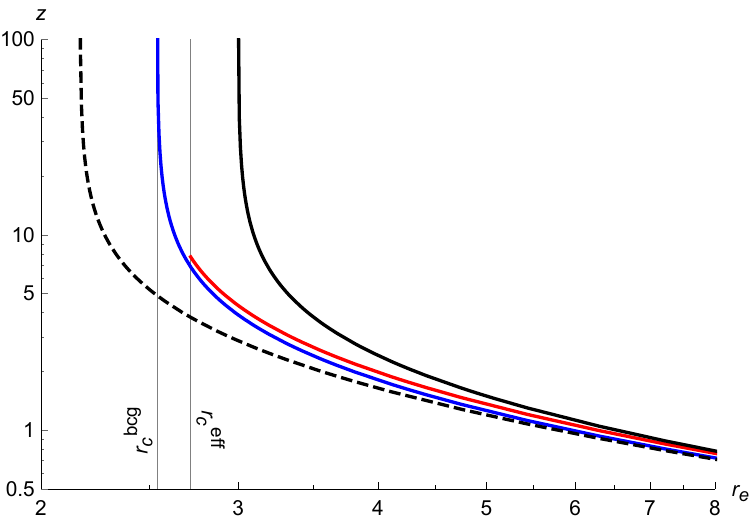}
	\caption{\textbf{Gravitational redshift: retrograde emission.} We display the gravitational redshift of a photon emitted in the direction opposite to the direction of motion of a massive test particle on a circular orbit, as observed by a static observer at infinity, as a function of emitter's position $r_e$. The legend and the parameters of black holes are identical to Fig.~\ref{fig:gredshift2}. }
\label{fig:gredshift3}
\end{center}
\end{figure}

For 
massive test particles on a circular trajectory, the conserved quantities $E_i=-g_{tt}(r_i)U_i^{t}$ and $L_i=g_{\varphi\varphi}(r_i)U_i^{\varphi}$ can be calculated from the conditions on circular timelike geodesics as in \eqref{eq:Ecircular} and \eqref{eq:Lcircular} with $\kappa=-1$ (thus $q_{\mu\nu}=g_{\mu\nu}$) and replacing $r_c$ by $r_i$. This gives the following relation:
\be
\Omega_i=-\frac{g_{tt}(r_i)L_i}{E_i g_{\varphi\varphi}(r_i)}=\pm\sqrt{\frac{f'(r_i)}{2r_i}}\,.
\ee
Also, for $U_i^t$, upon using \eqref{eq:Ecircular}, we get 
\be
U_i^t =\frac{E_i}{-g_{tt}(r_i)}= \frac{-1}{g_{tt}(r_i)}\lr{.}{\sqrt{\frac{-2g_{tt}^2}{2g_{tt}-rg_{tt}'}}}{|}_{r=r_i}=
\frac{1}{\sqrt{H(r_i)}}\,,
\ee
where 
\be
H(r_i)\equiv f(r_i)-\frac{r_i}{2}f'(r_i)\,.
\ee
Plugging this in \ref{eq: redshift not yet}, we finally get
\begin{align}\label{eq: redshift}
    z=&\sqrt{\frac{H(r_o)}{H(r_e)}}\lr{(}{\frac{-1 + b_\gamma\Omega_e}{-1+ b_\gamma\Omega_o}}{)}-1\,.
\end{align}
This formula gives the total redshift of a light ray emitted tangentially from a timelike geodesic at a point where its radial velocity vanishes. If the sign of $b_\gamma$ matches that of $\Omega_e$, the light is emitted in the direction of the orbiting massive particle. Moreover, if the sign of $b_\gamma$ matches that of $\Omega_o$, the light is observed as approaching the observing test particle in its direction. Thus $z$ given by \eqref{eq: redshift} accounts for the gravitational redshift and relativistic Doppler effect depending on the signs of $b_\gamma$, $\Omega_e$ and $\Omega_o$.\\
The dependence of $z$ on $r_e$ cannot continue below the \emph{background metric} photon sphere of the black hole given by $r_c$ in \eqref{eq:rc}, since its position is given by a root of the function $H(r)$, where the energy and angular momentum of the emitting test particle diverge as per \eqref{eq:Ecircular} and \eqref{eq:Lcircular}.\\
In the case of a distant observer $r_o\to\infty$ we get naturally $\Omega_o\to0$ and $H(r_o)\to1$ since $H(r)=1-M/r+O(1/r^3)$. So the formula then reduces to
\begin{align}\label{eq:redshifttangent}
z=\frac{1-b_\gamma\Omega_e}{\sqrt{H(r_e)}}-1.
\end{align}
A plot of the frequency shift in the distant observer scenario where the light ray is sent \emph{in}/\emph{against} the direction of the test particle is shown in Fig.~\ref{fig:gredshift2}/Fig.~\ref{fig:gredshift3}, respectively.

In fact, there is the following subtlety. 
The redshift $z=-1$ in 
Fig.~\ref{fig:gredshift2} corresponds to an ``infinite blueshift'': photons emitted with infinitely small frequency are observed at infinity with a finite frequency. 
In all cases except for the effective metric this happens precisely on the photon sphere where the energy of a timelike test particle on a circular orbit diverges due to it approaching the (background metric) speed of light. However, in the effective metric, the photon with appropriate polarization has timelike propagation vector $k^\mu$ and effectively moves slower than `regular' (background metric) light in this direction on this radius,  as is evident from the photon phase velocities in the $\varphi$ direction given by \eqref{eq: phase velocity phi eff} and \eqref{eq: phase velocity phi bcg}.
At some radius $r_s$, greater than $r_c^{\mbox{\tiny eff}}$,  the emitter's velocity precisely equals the photon's in effective metric -- photons emitted with  vanishing frequency, $\omega_e\approx 0$, will be measured with finite frequency $\omega_o$ further away  -- we have an infinite relative blueshift $z=-1$. The position of $r_s$ is simply given by $1=(b_\gamma\Omega_e)_{r=r_s}$, or 
\be
\Bigl(\frac{rf'}{2fh}-1\Bigr)_{r=r_s}=0\,.
\ee
Below this radius the frequency of light measured by the emitter, defined as \lq oscillations in the electromagnetic field per unit time\rq, starts increasing again
as the emitter velocity is required to increase.
However $\omega_e$ as defined by \eqref{eq: frequency} measures \lq phase frequency\rq~of the plane wave and since the wave-phase is decreasing as the emitter below $r_s$ moves past it, $\omega_e$ turns negative. Thus, Fig. \ref{fig:gredshift2} displays $z$ with regards to how the emitter \lq experiences\rq~the light, with $1+z=\abs{\omega_e}/\omega_o$.\footnote{Note that $z$ would in principle diverge, as $r$ diminishes towards $r_c^{\mbox{\tiny bcg}}$. However, photons emitted below $r_c^{\mbox{\tiny eff}}$ will never reach infinity, and for this reason, we truncate the red curve at $r_c^{\mbox{\tiny eff}}$.}

\begin{figure}
\begin{center}
	\includegraphics[scale=0.67]{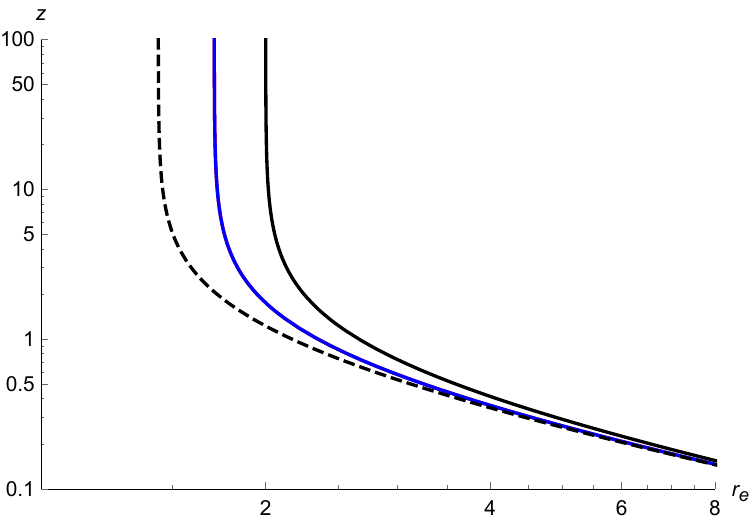}
	\caption{\textbf{Gravitational redshift: radial motion.} This picture shows the gravitational redshifts $z$ of a photon emitted radially outward and observed at infinity, displayed as function of static emitter position $r_e$ for the two RegMax metrics (shown in blue), fixing  $M=1$, $\alpha=1/\sqrt{2}$ and $Q=0.9$. These two curves coincide, as radial geodesics are not affected by the birefringence (cf.  Fig.~\ref{fig:phasevel}). For comparison we also show the Schwarzschild case ($M=1$) in solid black, and the Reissner--Nordstr\"om case ($M=1$, $Q=0.9$) in dashed black. The vertical asymptotes mark the black hole horizons, as expected.
    }
\label{fig:staticgredshift}
\end{center}
\end{figure}

In the case of a static emitter the redshift follows directly from the definition of $z$, the Killing symmetry conserving $E_\gamma=-k_t=-q_{tt}k^t$ and the normalization of four-velocity, due to which it holds that $U^t=\sqrt{\frac{-1}{g_{tt}}}$. Then:
\begin{align}
    1+z=\frac{\omega_e}{\omega_o}=\frac{q_{tt}k^tU^t|_e}{q_{tt}k^tU^t|_o}=\frac{U_e^t}{U_o^t}=\sqrt{\frac{g_{tt}(r_o)}{g_{tt}(r_e)}}\,.
\end{align}
Note that this is independent of whether the light travels w.r.t. the background or effective metric and it only depends on the four-velocity of the static test particles; this is consistent with the conclusion that for radial geodesics there is no birefringence. For a distant observer this formula reduces to
\begin{align}
    z=\sqrt{\frac{1}{-g_{tt}(r_e)}}-1.
\end{align}
Static redshift for RegMax black holes is 
shown in Fig. \ref{fig:staticgredshift}, where it is also compared to the Schwarzschild and Reissner--Nordström cases.

\subsection{Black hole shadow}
Each black hole casts a shadow. If the black hole is non-rotating, its shadow is circular and is produced by photons which escape from their unstable circular orbit to infinity due to small perturbations. Because of this fact, shadow radius coincides with the critical impact parameter which is a quantity with dimensions of length interpreted as the radial distance of the asymptotic photon trajectory from the scattering centre and, in accordance with \eqref{def null impact parameter}, it is in our simplified case 
(of spherically symmetric asymptotically flat spacetime and distant observer) given by \cite{Perlick:2021aok}
\begin{align}
r_{sh}=b_c=\sqrt{-\frac{r_c^2}{q_{tt}(r_c)}}.\label{bcrit}
\end{align}
For our two metrics, we can write explicitly
\begin{align}
r_{sh1}=\sqrt{\frac{r_c^2}{f(r_c)h(r_c)}}\,,\label{b1}
\end{align}
and
\begin{align}
r_{sh2}=\sqrt{\frac{r_c^2}{f(r_c)}}\,.\label{b2}
\end{align}

\begin{figure}
\begin{center}
	\includegraphics[scale=0.67]{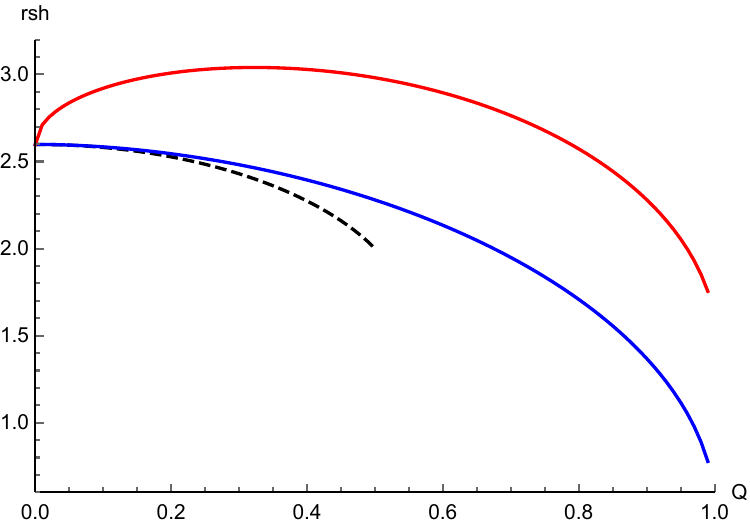}
	\caption{\textcolor{black}{\textbf{Black hole shadow: comparison.} We display the shadow radius as a function of electric charge $Q$, for various black holes with $M=1/2$ (and $\alpha=0.75$). Namely, the results for background (blue line) and effective (red line) RegMax metrics are compared to the  Reissner-Nordström case (dashed black line). On a given RegMax curve, to the left, we start with the `Schwarzschild-like' branch of black holes. As $Q$ increases, there is a transition to the Reissner--Nordstrom-like  branch, which eventually terminates at $Q$ corresponding to an extremal black hole. 
 }}
\label{fig:rsh}
\end{center}
\end{figure}

\begin{figure}
\begin{center}
	\includegraphics[scale=0.67]{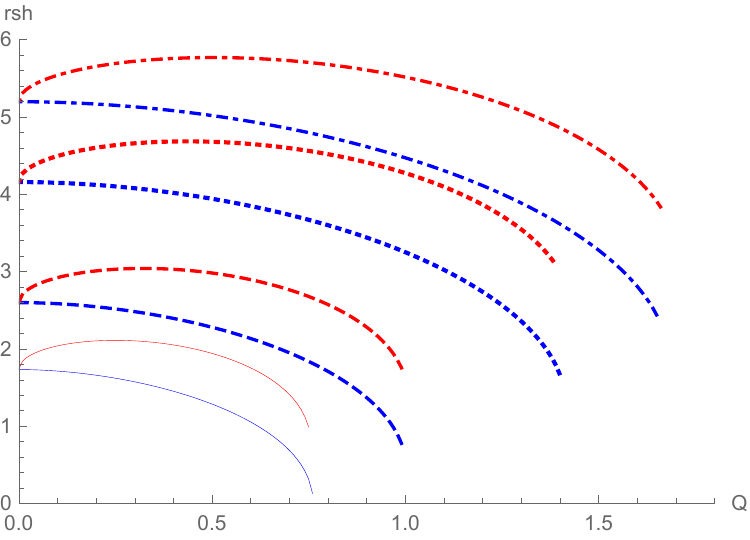}
	\caption{\textbf{Black hole shadow: mass dependence.} In this picture, we display the radii of the black hole shadows as functions of black hole charge for various black hole masses, and fixed $\alpha = 0.75$.  Results for the background metric are represented by blue curves, results for the effective metric by red curves. From bottom to top, masses are $M=1/3$ (thin curves), $M=1/2$ (dashed), $M=4/5$ (dotted), and $M=1$ (dot-dashed). Similar to the previous figure, charge $Q$ goes from zero to its maximal value corresponding to extremal black hole.}
\label{fig:shadowmass}
\end{center}
\end{figure}

We need to be careful though, since the photon sphere radius $r_c$ 
is different for the effective and background metrics. In the latter case, it can be expressed explicitly as shown in Eq. \eqref{eq:rc}, however, for the effective metric, it must be calculated numerically. In Fig. \ref{fig:rsh}, we can see the results for the RegMax theory compared with Reissner-Nordström case.
The dependence of the shadow radii on the black hole mass for RegMax black holes is displayed in Fig.~\ref{fig:shadowmass}.

\section{Lyapunov exponents and thermodynamic phase transitions}\label{sec:TDs}
As shown in \cite{Kubiznak:2023emm}, the RegMax AdS black holes feature rather interesting thermodynamic behavior and phase structure, including the existence of critical points, first order-phase transitions, and so on.
As conjectured in 
\cite{Zhang:2019tzi, Guo:2022kio, Yang:2023hci, Lyu:2023sih, Kumara:2024obd}, some of this behavior should be encoded in the corresponding behavior of the Lyapunov exponents. In this section, after reviewing the main thermodynamic features of RegMax black holes, we shall test this hypothesis. 

Importantly, the RegMax black holes provide an ideal testground to test this conjecture for two reasons. i) The interesting thermodynamic behavior prevails in both the canonical (fixed charge) and the grand canonical (fixed potential) ensembles, allowing us to test whether the Lyapunov exponents can know about/distinguish the two cases. ii) Due to birefringence, one has two kinds of Lyapunov exponents -- one associated with the background metric and another with the effective metric. Is the information about thermodynamics encoded in both of these?
In this section we focus on the AdS case, $\Lambda<0$. We start with the canonical ensemble.

\subsection{Canonical ensemble}

In the canonical ensemble, the free energy 
\be 
F=M-TS=F(T,Q,P,\alpha)
\ee 
is the key quantity. The global minimum 
of $F$ corresponds to the equilibrium phase; its non-analyticity indicates the presence of various phase transitions. For RegMax black holes the behavior of $F$ was studied in \cite{Kubiznak:2023emm}. We shall first review these results and then investigate whether they can also be reproduced by studying the Lyapunov exponents.

It turns out that the thermodynamic behaviour of RegMax black holes is characterized by two regimes, \lq Schwarzschild-like\rq~and \lq Reissner--Nordstr\"om-like\rq, depending on the value of parameter $\alpha$. Namely, there exists a `critical'  $\alpha$, 
\be \label{alphac}
\alpha_c=\frac{1}{\sqrt{2\abs{Q}}}\,,
\ee
independent of $P$, that separates these two regimes.
Such $\alpha_c$ can be determined \cite{Kubiznak:2023emm} from the behavior of the metric function $f$ near the origin, and separates the corresponding Schwarzschild-like and Reissner--Nordstrom-like horizon structure of RegMax black holes (cf.  Fig.~\ref{fig:fm}). Naturally, the horizon structure is also reflected by the corresponding thermodynamics.

\begin{figure}
\begin{center}
	\includegraphics[scale=0.67]{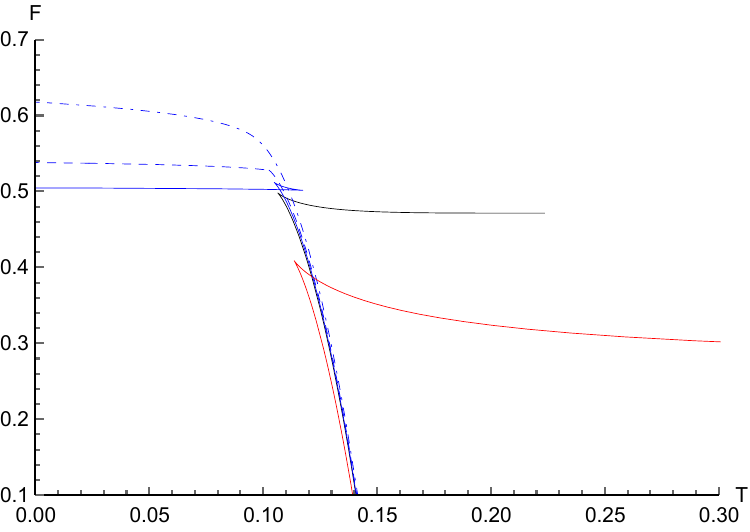}
	\caption{\textbf{$F-T$ diagram:} Free energy of the RegMax black hole is displayed for fixed $Q=1$ and fixed $P=0.023$. Depending on the value of $\alpha$, we  observe two Reissner--Nordstr\"om-like (blue) and Schwarzschild-like (red) regimes. In the red regime, represented by a red curve for 
 $\alpha=0.4<\alpha_c$. the free energy features a temperature gap, two branches of black holes, and a simple cusp; there is only one stable black hole phase corresponding to the lower branch of large black holes.  The marginal case,  $\alpha=\alpha_c$, is displayed by a black curve, for which the upper branch of (small) black holes has a termination point. Slightly above $\alpha_c$, we have two stable black hole phases (of large and small black holes) separated by a first-order phase transition, associated with the characteristic swallow tail behavior shown by a solid blue curve for $\alpha=0.76$. As $\alpha$ increases further, the swallow tail diminishes, and eventually terminates at a critical point,  where a second-order phase transition occurs, see blue dashed curve for $\alpha=0.82$.  For even higher values of $\alpha$, one cannot distinguish between the two phases and the free energy is smooth, as shown by a blue dot-dashed curve for $\alpha=1$. 
}\label{fig:FT_diagram}
\end{center}
\end{figure}

Namely, for 
$\alpha>\alpha_c$ we get the \lq Reissner--Nordstr\"om\rq~regime, known from the study of charged-AdS black holes in Maxwell’s theory \cite{Kubiznak:2012wp}, with a characteristic swallow-tail behavior of the free energy and the corresponding Van der Waals-like $P-T$ phase diagrams. This regime features a first-order phase transition between small and large black holes and the associated critical point where the first-order phase transition coexistence line terminates and the phase transition becomes of the second-order. On the other hand, for $\alpha<\alpha_c$, the thermodynamic behavior is Schwarzschild-like and features no black hole regions and regions with large stable black holes. The detailed discussion of these (and the marginal, $\alpha=\alpha_c$) cases, including the corresponding free energy and $P-T$ phase diagrams diagrams, can be found in  \cite{Kubiznak:2023emm}.

To capture all of the above thermodynamic features, we can simply probe the thermodynamic phase space, fixing the charge $Q=1$ and the thermodynamic pressure $P\approx 0.023$  (corresponding to $\ell$=2.3), and varying the parameter $\alpha$. We display the corresponding $F-T$ diagram in Fig.~\ref{fig:FT_diagram}. Here, the Reissner--Nordstr\"om regime is denoted by blue curves, Schwarzschild regime by red curve, and marginal case by black curve.\footnote{Alternatively, for fixed $\alpha$, Eq.~\eqref{alphac} determines critical charge $Q_c$. Thus, instead of fixing $Q$ and varying $\alpha$, one could equally fix $\alpha$ and vary $Q$. $Q>Q_c$ would then correspond to the Schwarzschild-like regime, and $Q<Q_c$ to the Reissner--Nordstr\"om-like regime. Since varying $Q$ seems  more `physical' than varying $\alpha$, one could then also construct the corresponding $Q-T$ phase diagram, similar to what was done in \cite{Chamblin:1999tk} for the Maxwell case.} 
Namely, for $\alpha=0.76$ (solid blue curve), slightly above $\alpha_c$, we observe the characteristic swallow tail 
behavior of the free energy, associated with the small black hole/large black hole phase transition.  
The choice $\alpha=0.82$ corresponds to the critical point, where the swallowtail degenerates into a single point. Beyond the critical point (for even larger $\alpha$) the $F-T$ curve becomes smooth and we can no longer distinguish large from small black holes.

On the other hand, for $\alpha<\alpha_c$ (red regime), the $F-T$ diagram features two branches of black holes meeting at a cusp. The lower branch corresponds to a stable phase of large black holes, while the upper (small black hole) branch is unstable. We also have a no black hole region, associated with the temperature gap in the $F-T$ diagram. Let us finally notice the special termination point for the upper branch of small black holes that occurs for the marginal,    $\alpha=\alpha_c$, case. Its temperature coincides with the finite asymptotic temperature in the corresponding $P-T$ phase diagram, see \cite{Kubiznak:2023emm}.

\begin{figure}
\begin{center}
	\includegraphics[scale=0.67]{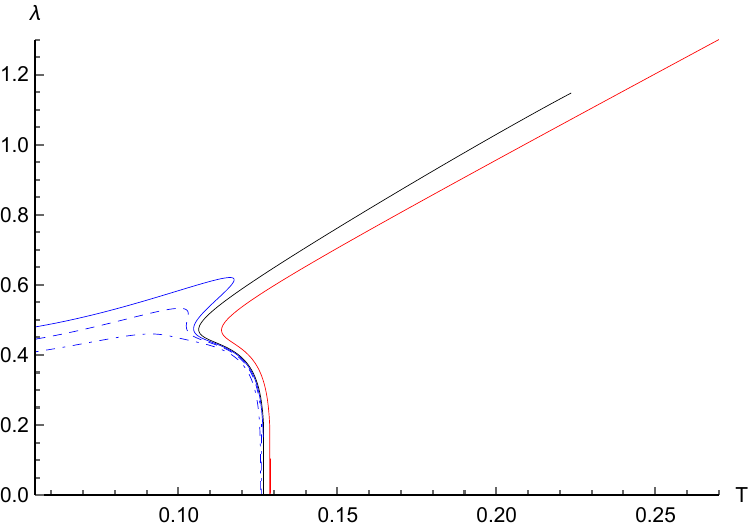}
	\caption{\textbf{$\lambda-T$ diagram: timelike geodesics.} Displayed  curves correspond to $\alpha=0.4$, $\alpha=1/\sqrt{2}$, $\alpha=0.76$,  $\alpha=0.82$, and $\alpha=1$, from right to left. Cusps in the $F-T$ diagram are here represented by single bends, the presence of first-order phase transitions by `double-bended' curves and the second-order phase transition by an inflection point in the $T=T(\lambda)$ behavior. Note that the quickly dropping branch (of large black holes) terminates at $\lambda=0$, corresponding to finite $r_+$. We have set $L=50$, and, as in Fig. \ref{fig:FT_diagram}, $Q=1$ and $P=0.023$.
}\label{fig:lambdat_tl_diagram}
\end{center}
\end{figure}

{\em Timelike geodesics.}
The crucial question now is: are we able to restore these results and find phase transitions using Lyapunov exponents? To that end, we will study $\lambda-T$ diagrams, starting with the case of massive particles displayed in Fig.~\ref{fig:lambdat_tl_diagram}. 
To plot these, we 
first numerically determine $r_c=r_c(r_+,Q,\alpha,P, L)$  from Eq.~\eqref{eq:Lcircular}. This then allows one to express 
$
\lambda=\lambda(r_+,Q,\alpha, P, L)\,,
$
employing 
Eq.~\eqref{Lambda_timelike}, 
and since we know $T=T(r_+,Q,\alpha,l)$, \eqref{Tform}, we effectively have 
\be
\lambda=\lambda(T,Q,\alpha, P, L)\,,
\ee
which can be plot parametrically, fixing $Q,\alpha, P$, and $L$, and varying $r_+$.  

The curves {in the $\lambda-T$ diagram in  Fig.~\ref{fig:lambdat_tl_diagram} have the same colouring and values of $Q, P, \alpha$, as those in the $F-T$ diagram in Fig. \ref{fig:FT_diagram}.
For $\alpha = 0.4$, there is a cusp in the $F-T$ diagram -- this maps onto the 
`red curve bend' in the $\lambda-T$ diagram. The branch of large black holes corresponds to the quickly dropping to $\lambda\approx 0$ 
branch in the $\lambda-T$ diagram.} Since the upper red branch (of small black holes) in the $F-T$ diagram goes to infinity, also the red curve in $\lambda-T$ diagram extends to arbitrary large temperatures. We get a similar feature also for $\alpha=\alpha_c$ (black curve) but since its $F-T$ diagram has an endpoint from where small black hole branch emerges, also its $\lambda-T$ diagram ends at the corresponding finite temperature.

For $\alpha=0.76$, we observe a swallowtail in $F-T$ diagram. In the $\lambda-T$ diagram, it corresponds to a blue curve with `two bends' whose temperatures match with temperatures of outer points of the swallowtail. At $\alpha\approx0.82$, represented in $\lambda-T$ diagram by a blue dashed curve, 
the second-order phase transition occurs. It corresponds to  
an inflection point in the $T=T(\lambda)$ behavior, determined from 
\be \label{infllambda}
\pdv{T}{\lambda}=0=\pdv[2]{T}{\lambda}\,.
\ee
The temperature of this inflection point is the corresponding temperature of the thermodynamic  critical point. Finally, for $\alpha>0.82$, $F-T$ diagram is smooth without any features and so is the $\lambda-T$ diagram.

\begin{figure}
\begin{center}
	\includegraphics[scale=0.67]{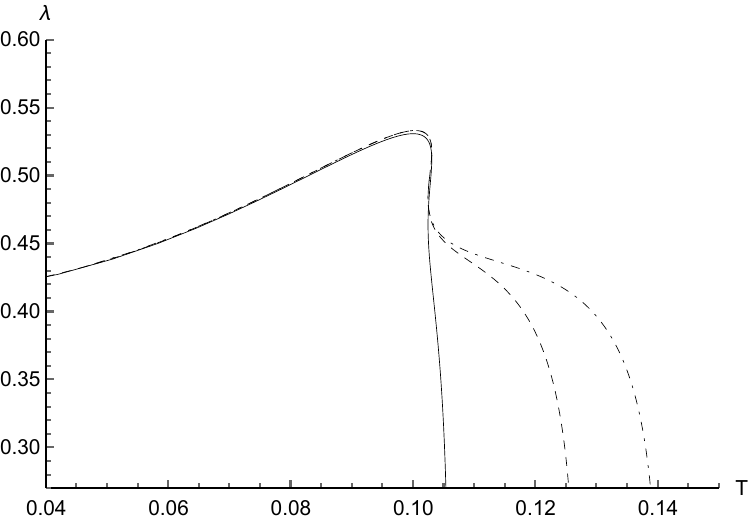}
	\caption{\textbf{Effect of $L$ on $\lambda - T$ diagram.} Here, we display the effect of various choices of $L$ on the shape of $\lambda - T$ diagram for a representative (critical point) case of  $\alpha\approx0.82$. Namely, $L=10$ corresponds to solid curve, $L=50$ to dashed curve, and $L=100$ to dot-dashed curve. As obvious, the critical temperature remains the same irrespective of which $L$ we choose; the choice of $L$ only affects where our curves finally `go down'.
}\label{fig:diferL}
\end{center}
\end{figure}

\begin{figure}
\begin{center}
	\includegraphics[scale=0.67]{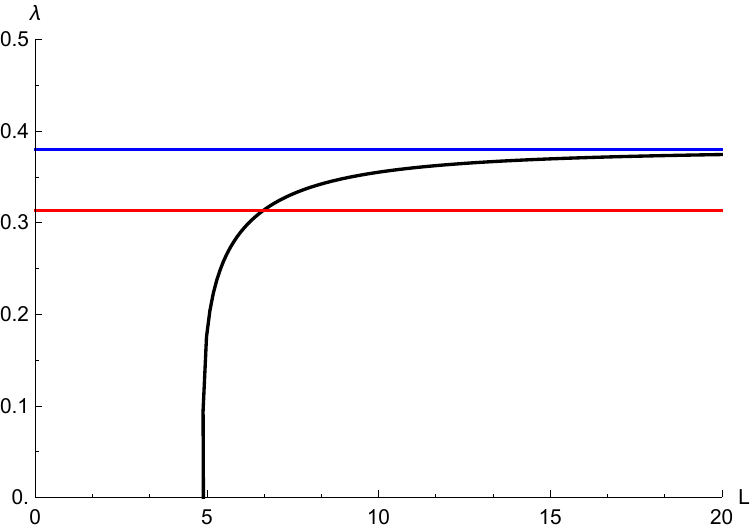}
	\caption{\textbf{Critical point Lyapunov exponent: dependence on $L$.} For timelike geodesics (black curve), the value of the Lyapunov characteristic exponent $\lambda$ 
 at the critical point, 
 $T=T_c$, depends on the angular momentum $L$. In particular,  note that an unstable circular timelike geodesic only exists for particles with $L$ greater than some minimum value, cf. the behavior of the corresponding effective potential in Fig.~\ref{fig:V-L}. The dependence on $L$ of course disappears for null 
    effective (red curve) or null background (blue curve) geodesics, cf. Fig.~\ref{fig:V-L-null}. 
 The plot is shown for values $Q=1, \alpha\approx1$, and $P=P_c\approx 0.01$.
}\label{fig:lambdaL}
\end{center}
\end{figure}

One can be concerned about the effect of angular momentum $L$ on the shape of the above curves. {Nonetheless, it turns out that $L$ only influences `how quickly' the curves `go down' to $\lambda=0$, but the choice of $L$ (as long as it is sufficiently large) has no impact on the features connected with phase transitions, as  shown in Figs.~\ref{fig:diferL} and \ref{fig:lambdaL}. In particular, the critical point in RegMax depends on the choice of $Q$ and $\alpha$, which uniquely determine the critical black hole radius $r_c$, see Eq.~\eqref{rccccc} below, the critical temperature $T_c$, and the critical pressure $P_c$. In the $\lambda-T$ diagram, it corresponds to the inflection of $T=T(\lambda)$ in the blue dashed curve, determined from \eqref{infllambda}. This point always lies on $T=T_c$, however the value of $\lambda$ is not given by the black hole thermodynamics,  and depends on the angular momentum $L$ of the test particle. For a particular case, such a dependence is shown in Fig.~\ref{fig:lambdaL}. We see that 
for timelike geodesics there exists a minimal value of $L$,  
below which there are no unstable circular geodesics, cf. Fig.~\ref{fig:V-L}. Of course, for both effective null (red) and background null (blue) geodesics, the critical value of $\lambda$ is independent of $L$, as displayed in Fig.~\ref{fig:lambdaL} (see also Fig.~\ref{fig:V-L-null}).  
}

\begin{figure}
\begin{center}
    \includegraphics[scale=0.67]{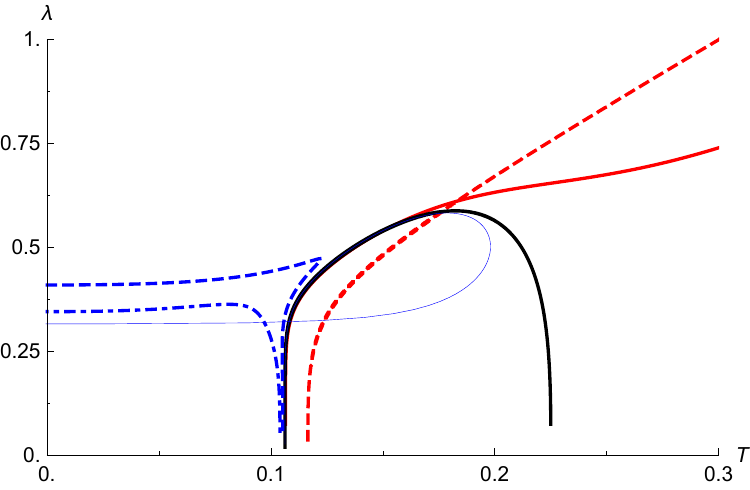}
	\caption{\textbf{$\lambda-T$ diagram: effective null geodesics.} While the $\lambda-T$ dependence for null effective geodesics seems to capture all interesting thermodynamic behavior, it is rather complicated. In particular, for $\alpha=\alpha_c$, we develop a `hook' (black curve) and for slightly higher $\alpha$, there is a self-intersection (solid blue curve). In this plot, we have set  $Q=1, l=2.3$, and displayed  $\alpha=0.5$ (dashed red), $\alpha=0.7$ (solid red), $\alpha=1/\sqrt{2}$ (black), $\alpha=0.708$ (solid blue), $\alpha=0.75$ (thin blue), $\alpha=0.822$ (dashed blue) and $\alpha=1$ (dot-dashed blue). Although it is not  obvious, solid and dashed blue curves display both cusps of the swallow tail; the corresponding first-order phase transition temperature is, however, greater than $T(\lambda=0)$.  
}\label{fig:efflambda-T}
\end{center}
\end{figure}

{\em Null effective geodesics.} Let us next turn to the null geodesics in the effective metric. The corresponding $\lambda-T$ diagram
for different values of $\alpha$
is displayed in Fig.~\ref{fig:efflambda-T}. In this case, the Lyapunov exponent is given by \eqref{lambda1} where, however, now $r_c$ is determined from \eqref{null derivative reduction}, and it is manifestly independent of $L$:
\be\label{lambda-plott}
\lambda_{\gamma}=\lambda_\gamma(T,Q,\alpha, P)\,.
\ee
Although the $\lambda-T$ dependence seems to display all interesting thermodynamic features, 
it is now rather complicated.  
In particular, for $\alpha=\alpha_c$ the dependence develops a \lq hook\rq  (solid black curve). This causes the $\lambda-T$ curve for $\alpha>\alpha_c$, $\alpha$ very close to $\alpha_c$, to self intersect. Such intersection, however, does not correspond to the `bottom of the swallow tail' and the associated first-order phase transition is not located in that intersection, although the latter is present whenever 
 $\lambda(T)$ is triple valued, as is the case of solid blue and dashed blue curves.
Note also that, as $r_+$ increases, the curves quickly fall to $\lambda=0$, similar to 
what happens for timelike geodesics, only now $\lambda$ is independent of $L$.\footnote{While this is not the case with the present figures, it remains to be seen whether the restriction $\lambda\geq 0$, translated to the restriction on how large $r_+$ can be captured by $\lambda-T$ timelike or null effective diagrams, can effectively erase some important thermodynamic features (that would have been present for large enough black holes). In fact, the first-order phase transition for the blue curves displayed in Fig.~\ref{fig:efflambda-T}, occurs for temperatures greater than $T(\lambda=0)$. }

\begin{figure}
\begin{center}
	\includegraphics[scale=0.67]{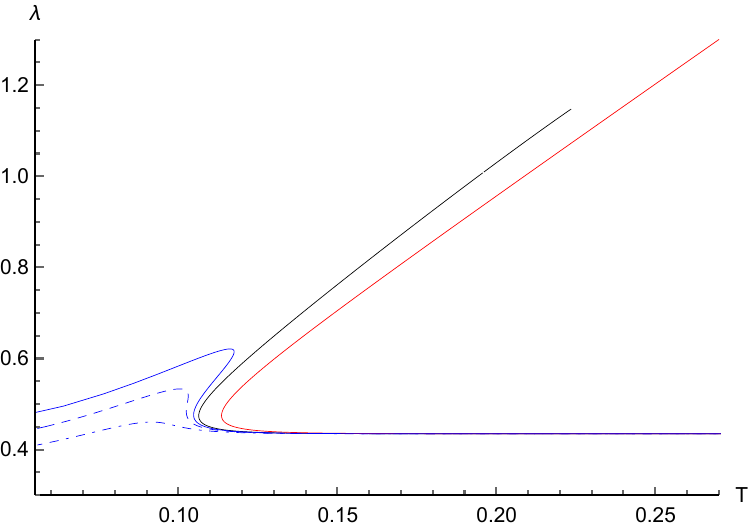}
	\caption{\textbf{$\lambda-T$ diagram: null background geodesics. } The values of $\alpha's$ and coloring are the same as in Fig \ref{fig:lambdat_tl_diagram}. 
 Since (for large black holes) the curves do not `go down' but continue to large temperatures, in this case we can clearly identify all thermodynamic features, including the critical point (inflection), first order phase transition (triple valued curves), and the cusp behavior. 
 }\label{fig:lambdat_null_diagram}
\end{center}
\end{figure}

{\em Null background geodesics.}
Finally, we study the Lyapunov exponents for the null geodesics in the background metric, as displayed in Fig.~\ref{fig:lambdat_null_diagram}. It corresponds to $\lambda$, given by 
\eqref{lambda2} where, again, $r_c$ is determined from \eqref{null derivative reduction}, and it is manifestly independent of $L$, namely formally we get again the expression \eqref{lambda-plott}.  The resultant $\lambda-T$ diagram looks `significantly simpler' than in the effective or timelike cases. In particular, 
for large black holes $\lambda$ continues parallel with the $T$ axis without collapsing to zero at some specific point, as it happens for timelike or null effective geodesics. 
The presence of critical point, first-order phase transition, or the cusp are now apparent from this figure.

\subsection{Grand-canonical ensemble}
\begin{figure}
\begin{center}
	\includegraphics[scale=0.77]{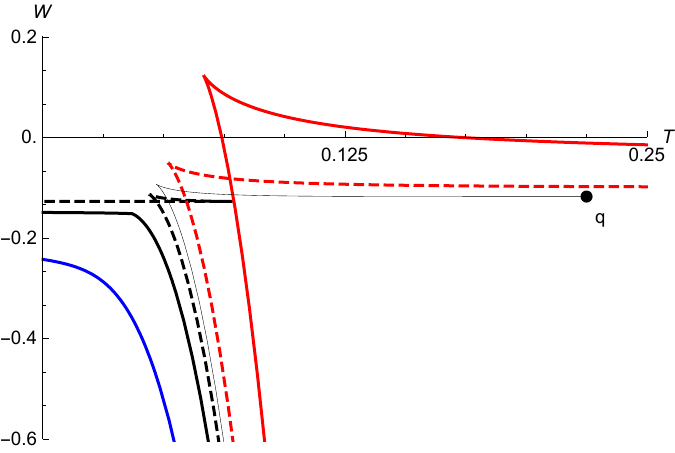}
	\caption{\textbf{$W-T$ diagram:} Here, we display the grand-canonical free energy $W$ as a function of temperature $T$ for $\alpha=1$ and $P=0.01$, and 
 various values of $\phi$: $\phi=0.5$ (solid red), $\phi=0.67$ (dashed red), $\phi=\frac{1}{\sqrt{2}}$ (thin black), $\phi=0.725$ (dashed black), $\phi=0.76$ (solid black) and $0.85$ (solid blue). Similar to the canonical ensemble, the diagram features cusps, swallow tails, termination point, or a critical point. When interpreting the corresponding phase diagram, however, one also has to take into account the possibility of the radiation phase, characterized by $W\approx 0$, see \cite{Kubiznak:2023emm} for details.} 
\label{fig:WTdiagram}
\end{center}
\end{figure}

\begin{figure}
\begin{center}
	\includegraphics[scale=0.67]{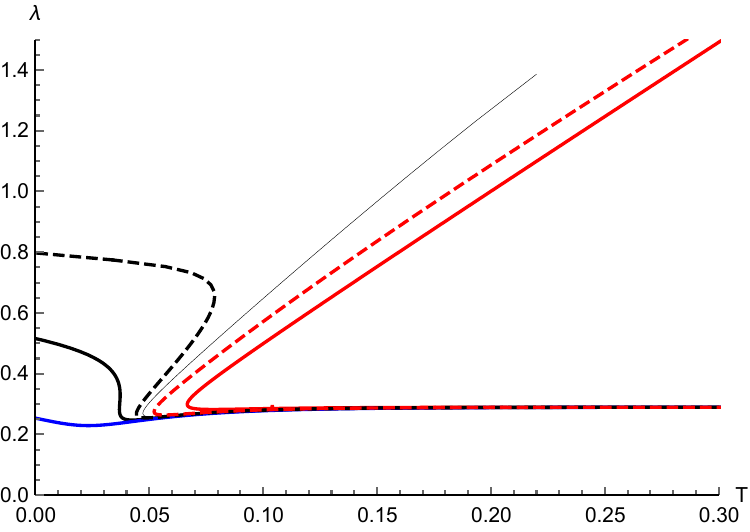}
	\caption{\textbf{Grand-canonical ensemble: $\lambda-T$ diagram for background null geodesics.} The figure is displayed for the same values of parameters and with the same colouring as in Fig.~\ref{fig:WTdiagram}. We clearly see the presence of the critical point (solid black), as well as the 1st-order phase transition (triple valued dashed black), and minimal temperature cusps. 
            }
\label{fig:GCphieffect}
\end{center}
\end{figure}

\begin{figure}
\begin{center}
    \includegraphics[scale=0.67]{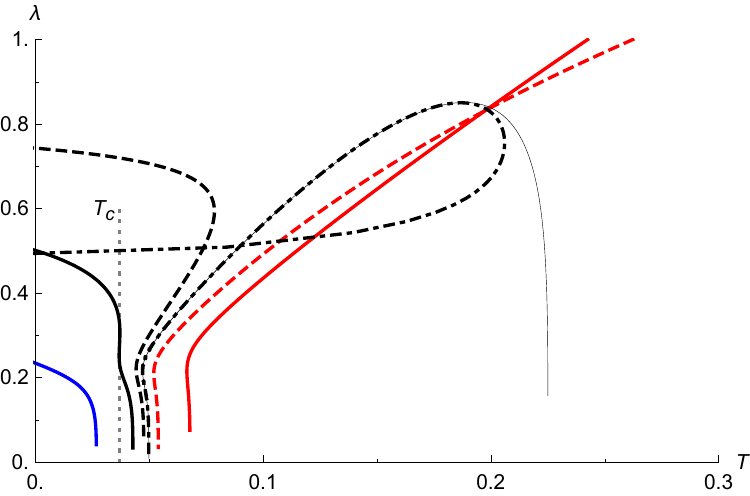}
	\caption{\textbf{Grand-canonical ensemble: $\lambda-T$ diagram for effective null geodesics. } The values of parameters and the colouring is the same as in Fig.~\ref{fig:WTdiagram} except for an additional dot-dashed black line with $\phi=0.7072$ to show the intersection behaviour near $\phi_c$.
            }
\label{fig:GCphieffecteff}
\end{center}
\end{figure}

The RegMax model has also been shown to have interesting thermodynamic behaviour in the grand canonical (fixed electric potential $\phi$) ensemble \cite{Kubiznak:2023emm}. In the following, we investigate whether this can also be linked to the Lyapunov exponents of unstable circular geodesics, similar to what happens in the canonical case.  In order to fix the potential $\phi$, we invert \eqref{phiTD}, and obtain
\begin{align}\label{Qphi}
Q=\frac{\phi}{2\alpha^2}\left(2\alpha^2r_++|\phi|+\sqrt{\phi^2+4|\phi|\alpha^2r_+}\right)\,. 
\end{align}
The potential governing the thermodynamic behavior is now the following (grand-canonical) free energy:
\be
W=M-TS-\phi Q=W(T,\phi, P, \alpha)\,,
\ee
whose global minimum determines the equilibrium phase. Note, however, that in this case, we have also a radiation phase, characterized by $W\approx 0$.
In contrast to the canonical case, where the phase behaviour switched between \lq Schwarzschild-like\rq~and \lq Reissner-Nordstr\"om-like\rq~for the critical $\alpha_c$, in the grand canonical case there exists a critical $\phi$: 
\be 
\phi_c=\frac{1}{\sqrt{2}}\,,
\ee
which causes the switch.\footnote{
To derive this, one simply plugs \eqref{Qphi} in formula \eqref{Tform} for
the temperature $T$, and expands for small $r_+$.}
In order to gauge the effect black hole phase transitions have on the Lyapunov exponents, we compare cases of $\phi$ corresponding to those discussed in Fig.~15 in \cite{Kubiznak:2023emm}. For reader's convenience, the associated $W-T$ diagram is reproduced in Fig. \ref{fig:WTdiagram}. Corresponding to this diagram we plot the Lyapunov exponents for null geodesics in background metric in Fig. \ref{fig:GCphieffect}, and null effective metric in Fig. \ref{fig:GCphieffecteff}.

As was the case with the canonical ensemble, the null background Lyapunov exponents can be used for phase transition analysis even here. A single bend in the $\lambda-T$ diagram (red curves in Fig.~\ref{fig:GCphieffect}) corresponds to a cusp in $W-T$ diagram, double bending is an indicator of the first-order phase transition (swallowtail in $W-T$ diagram) and the inflection point 
 of $T=T(\lambda)$ is related to the critical point of the second-order phase transition. Interestingly, similar structure is also observable (although in a bit more complicated manner) for the null effective Lyapunov exponents, displayed in Fig.~\ref{fig:GCphieffecteff}. Note in particular the presence of the `hook' and the corresponding `self-intersecting curve', similar to what happens in Fig.~\ref{fig:efflambda-T}.
 
\begin{figure}
\begin{center}
	\includegraphics[scale=0.67]{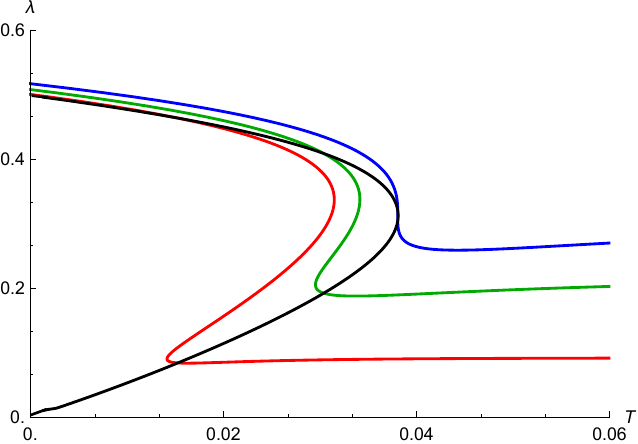}
	\caption{\textbf{Is there a Maxwell's equal area law?} Focusing on the null background metric $\lambda-T$ diagram, we display the first-order phase transition coexistence curve (shown in black) in the grand canonical ensemble corresponding to $\phi=0.76, \alpha=1$, and  various pressures: $P=0.001$ (red curve), $P=0.0055$ (green), and $P\approx0.0111$ corresponding to the critical point (blue). The presence of some kind of Maxwell's equal area law for Lyapunov exponents does not seem obvious from this figure. 
            }
\label{fig:GCcoexistence}
\end{center}
\end{figure}

As discussed above, the triple valued curves indicate the presence of the swallow tail and the asscoiated first-order phase transition. However, 
 the $\lambda-T$ diagram does not seem to be governed by some sort of equal area law, as would be expected if we were able to interpret $\lambda$ as some kind  of thermodynamic potential. In order to show this we can construct the coexistence curve using the regular formalism of black hole thermodynamics as done in \cite{Kubiznak:2023emm} and plot the corresponding points in the $\lambda-T$ diagram. This is done in Fig. \ref{fig:GCcoexistence} for null geodesics in background metric in the grand canonical ensemble for $\phi=0.76$, $\alpha=1$ with regular $\lambda-T$ curves for three different values (the colourful lines) of the pressure $P$. Intersections of the colourful lines lie directly above each other (equal $T$) and mark the phase transition temperature for the particular $P$. The point of maximum temperature of the coexistence curve marks the critical point which lies on the \lq unfolding\rq~point of the $\lambda-T$ curve with $P=P_c\approx0.0111$ (blue curve). It is evident that the phase transitions do not follow an equal area law in such a diagram.

\subsection{Universality of critical point criterion}\label{sec:universality}
As we have seen above, the temperature of the critical point can be determined from the $\lambda-T$ diagrams -- it corresponds to the temperature of $T(\lambda)$ inflection points, characterized by \eqref{infllambda}. Let us briefly comment here as to why this is possible.

For fixed pressure, the critical point can be determined from the following two conditions: 
\begin{align}\label{CritPointcond}
    \pdv{T}{r_+}=0=\pdv[2]{T}{r_+}\,.
\end{align}
Solving the second equation for $r_+$ (third degree polynomial; we take the only real positive solution) gives the critical horizon radius $r_{+c}(\alpha,Q)$:
\begin{align}
    r_{+c}=&-\frac{\sqrt{| Q| }}{\alpha}\nonumber\\
    &+2^{2/3} \alpha^{1/3} |Q|^{7/6}\lr{(}{\sqrt{1-2 \alpha^{2} |Q|}-1}{)}^{-1/3}\nonumber\\
    &+2^{1/3}\alpha^{-1/3}|Q|^{5/6} \lr{(}{\sqrt{1-2 \alpha^{2} |Q|}-1}{)}^{1/3}\,.\label{rccccc}
\end{align}
The equation is solvable due to the fact that the third derivative of the metric function $f$ w.r.t. $r_+$ ($f^{(3)}(r_+)\propto T''(r_+)$) no longer contains the logarithmic term. $r_{+c}$ immediately gives the critical volume $V_c(\alpha,Q)$ by \eqref{Vform}. Solving the first equation for $\ell$ and substituting $r_+$ for $r_{+c}$ yields the critical AdS radius $\ell_c$ (a rather long expression), which gives the critical pressure by \eqref{Vform}. Inserting the solutions into $T$ finally determines the critical temperature $T_c$, fully characterising the critical point.

Consider now a quantity $\Phi$ (for example, the Lyapunov exponent $\lambda$, the circular radius $r_c$, or the ISCO radius) that is a function of the horizon radius $r_+$. Then, by using \eqref{CritPointcond}, we have 
\begin{align}
    \pdv{T}{\Phi}=\frac{\pdv{T}{r_+}}{\pdv{\Phi}{r_+}}\equiv\frac{\dot{T}}{\dot{\Phi}}=0\,,\\
    \pdv[2]{T}{\Phi}=\frac{\dot{\Phi}\ddot{T}-\dot{T}\ddot{\Phi}}{\dot{\Phi}^3}=0\,,
\end{align}
provided $\dot{\Phi}$ and $\ddot{\Phi}$ exist and $\dot{\Phi}$ is nonzero. Thence, the critical point conditions \eqref{CritPointcond} imply in particular the conditions \eqref{infllambda}, relevant for the $\lambda-T$ diagrams. 
The Lyapunov exponents are a convenient choice of scalar function $\Phi$, 
as there is an effort to measure them for nearly-bound geodesics of the photon ring of the M87 black hole, e.g. \cite{Johnson:2019ljv, Gralla:2020srx}.\footnote{In fact, the correspondence between the maximal Lyapunov characteristic coefficient $\lambda$ and black hole thermodynamics is deeper than the position of the critical point \cite{Zhang:2019tzi}. It enables one to connect physics that is associated with the horizon (such as black hole thermodynamics) with physics in the vicinity of the photon sphere. For example, since $-q_{tt}'(r_c)>0$ from \eqref{null derivative reduction}, and in our case $h>0$ and $h'>0$, we have  $\pdv{f}{r_c}>0$. Moreover, since $T>0$, i.e., $\pdv{f}{r_+}>0$, by chain rule we have $\pdv{r_c}{r_+}>0$, which implies a direct sign correspondence between $\dv{\Tilde{\lambda}}{r_+}$ and $\dv{\Tilde{\lambda}}{r_c}$, where $\Tilde{\lambda}$ can be for example the Lyapunov coefficient $\lambda$, the temperature $T$, or the Ruppeiner scalar curvature \cite{Cai:2021uov}. In particular, the Lyapunov exponents have been studied in this context, for example in   \cite{Lyu:2023sih,Yang:2023hci,Guo:2022kio}. 
}

\subsection{Summary}

The maximal Lyapunov exponents $\lambda$ of null and timelike geodesics 
can be used to expose features of black hole thermodynamics such as the critical point if $\lambda>0$ for $r_+$ large enough so that the feature in question is present. For timelike geodesics this is a restriction on a minimal angular momentum $L$ of the test particle, for null geodesics in effective metric this means the feature may not be possible to observe. For null background geodesics this imposes no restriction and, e.g., the critical point should always be present in the $\lambda-T$ diagram as an $T(\lambda)$ inflection point.
However this is to be expected simply by the critical point criterion, as dicussed in the previous subsection. 
Thus, it seems that the Lyapunov exponents of null geodesics in background metric are, out of the presented cases, the simplest ones to study the thermodynamic features of RegMax black holes.

A natural question is whether $\lambda$ can be thought of as some sort of thermodynamic potential and so if there is some analogue of an \lq equal area law\rq~that would govern the location of a phase transition. Unfortunately we were unable to find any sort of transformation of $\lambda$ so that information about the phase transitions could be obtained in such a way.

Finally, from our investigation it seems that one can use the same Lyapunov exponents (without performing a Legendre transformation) to encode the thermodynamic behavior of both canonical and grand-canonical ensembles. In this respect, the Lyapunov exponents are `simpler' objects than the corresponding thermodynamic potentials.

\section{Conclusions}\label{sec:theend}

In this paper, we have studied various optical properties of recently constructed RegMax black holes. These are particularly interesting because of the presence of  birefringence, a generic property of theories of NLE, due to which the two electromagnetic modes propagate along two distinct (effective and background) metrics.

Consequently, we have identified 
two distinct photon spheres (one for each metric) and thus also two different black hole shadow radii; for a given black hole mass the effective photon sphere lies outside the background photon sphere, cf. Fig.~\ref{fig:photon trajectories}.
Especially interesting is the behavior of redshift for effective metric modes observed in Fig.~\ref{fig:gredshift2}. It is related to the fact that circular observes in a certain region can `move faster' than the effective light modes, giving raise to a peculiar behavior displayed by the red curve in this figure. 
Nevertheless, the effects connected to birefringence can be observed only when the motion is not entirely restricted to radial direction, that is, the corresponding propagation vector has nontrivial angular components.
This is also why we should expect that standard thermodynamics (with Hawking temperature given by surface gravity) should remain valid for black holes in NLEs.

{Equally interesting is the thermodynamic analysis, exploiting the maximal Lyapunov exponent $\lambda$. Depending on the values of parameters of the theory, RegMax black holes feature Schwarzschild-like behaviour with no phase transitions or Reissner--Nordström-like behaviour, where first-order phase transition between small and large black holes is present and its coexistence curve terminates at a critical point.  

All these kinds of different behaviour could be identified from $\lambda - T$ diagrams for time-like as well as null geodesics. In this respect, the study of null geodesics in the background metric proved to be the most convenient choice. In particular, the critical point presents itself in the $\lambda-T$ curve in both the canonical and grand-canonical ensembles in a manner to be expected of any function of $r_+$, as discussed in the last section. Nonetheless,  we 
could not reproduce the entire coexistence curve -- an analogue of Maxwell's area law for Lyapunov exponents seems missing.

Finally, we hint that Lyapunov exponent is not the only scalar function which can be used for the purpose of thermodynamic analysis. In fact, the information about the critical point temperature is encoded in the event horizon radius $r_+$, so basically any function of the horizon radius can do the trick. The main selection criterion is therefore to choose such a function, which can be directly measured or easily calculated from experimental data, as is the case of Lyapunov exponents.

\subsection*{Acknowledgements}
We would like to thank Rick Perche, Ota Svitek, and Tay Tahamtan 
for useful discussions. 
D.K. and J.M. are grateful for support from GA{\v C}R 23-07457S grant of the Czech Science Foundation.

\appendix

\section{Derivation of the formula for Lyuapunov exponents}\label{A}
Here, following the 
path given in \cite{Cardoso:2008bp}, we re-derive the formula \eqref{def lambda} for Lyapunov exponents used in the main text. 
Let's start with the Lagrangian 
\be 
\mathcal{L}=\frac{1}{2}\left(q_{tt}\dot{t}^2 +q_{rr}\dot{r}^2+r^{2}\dot \varphi^2  \right) \,,
\ee
describing a geodesic motion in the equatorial plane, \eqref{Lagrangian of test particles}. Let $X_i=(p_q,q)_i$ denote a set of pairs of conjugate variables, obeying the following equations of motion:

\begin{align}
\dot{X_i}=F_i(X)\,, \label{eqm}
\end{align}
where 
the dot denotes the derivative with respect to `proper time' $\tau$.
By linearizing the above equations 
about some orbit of interest, ${\bar X}_i(\tau)$, we get 
\begin{align}
\delta \dot{X_i} (\tau)=K_{ij}(\tau)\delta X_j(\tau)\,, \label{linX}
\end{align}
with
\begin{align}
K_{ij}(\tau)=\dv{F_i}{X_j}\bigg|_{X={\bar X}(\tau)} \label{Kmatrix}
\end{align}
being the linear stability matrix. Its eigenvalues are called characteristic Lyapunov exponents; they are associated with a given point in phase space \cite{Springerlambda}. In the case of nonlinear systems, they can be used to distinguish stable and unstable equilibrium points. If the system is linear, the analysis is even simpler - eigenvalues of $K_{ij}$ characterize the stability of the whole system.

We can find solution of our linearized equation in terms of the evolution matrix $L_{ij}(\tau)$ as follow
\begin{align}
\delta X_i(\tau)=L_{ij}(\tau)\delta X_j(0)\,.
\end{align}
The evolution matrix must satisfy the following conditions:
\begin{align}
L_{ij}(0)&=\delta _{ij}\,, \label{L1}\\ 
\dot{L}_{ij}(\tau)&=K_{im}(\tau)L_{mj}(\tau)\,. \label{L2}
\end{align}
Eigenvalues of the evolution matrix lead to Lyapunov exponent
\begin{align}
\lambda=\lim_{\tau\to\infty}\frac{1}{\tau}\log{\left(\frac{\norm{\delta X(\tau)}}{\norm{\delta X(0)}}\right)}=\lim_{\tau\to\infty}\frac{1}{\tau}\log{\left(\frac{L_{ii}(\tau)}{L_{ii}(0)}\right)}\,. \label{Lyapunovexp}
\end{align}
which measures how much two nearby trajectories in phase space diverge: negative Lyapunov exponent means convergence whilst positive value indicates divergence of trajectories; zero Lyapunov exponent thence corresponds to stable, non-chaotic situation.
The linear stability matrix $K$ is constant when evaluated on the circular geodesic, so the equation \eqref{L2} has the formal solution
\be
L(\tau)=\exp(K\tau)\,,
\ee
which enables us to calculate Lyapunov exponents as eigenvalues of $K$.
Each eigenvalue corresponds to one $\lambda_i$ (which do not have to be distinct). These form the Lyapunov spectrum  \cite{Springerlambda2}. They can be ordered by size and the biggest one of them is known as maximal (or principal) Lyapunov exponent.

Now, let's calculate it explicitly for our case of interest. Since we are interested in radial motion, we take $X=(p_r,r)$. It is useful to recall the Euler--Lagrange equations and the formulas for canonical momenta:
\begin{align}
p_q&=\frac{\partial\mathcal{L}}{\partial \dot{q}}\,,\\
\dv{\tau}\left(\frac{\partial \mathcal{L}}{\partial \dot{q}}\right) &=\dv{p_q}{\tau}=\frac{\partial \mathcal{L}}{\partial q}\,.
\end{align}
Thus, we can see that for our $X=(p_r,r)$, we can rewrite Eq. \eqref{eqm} as
\begin{align}
\dot{p_r}&=\frac{\partial\mathcal{L}}{\partial r}\,, \label{pdot}\\
\dot{r}&=\frac{p_r}{q_{rr}} \label{rdot}\,.
\end{align}
Finally, if we want to express the quantities in coordinate time $t$ rather than the proper time $\tau$, 
we can make use of the relation
\begin{align}
\dv{\tau}=\dv{t}{\tau}\dv{t}=
\dot{t}\dv{t}\,,
\end{align}
which enables us to switch between proper and coordinate time derivatives.
If we use this and linearize \eqref{pdot},\eqref{rdot} about a trajectory with constant radius (considering Eqs. \eqref{linX} and \eqref{Kmatrix}), the corresponding linear stability matrix takes the following form: 
\begin{align}
K_{ij}=
\begin{pmatrix}
0 & K_1 \\
K_2 & 0 
\end{pmatrix}\,,
\end{align}
with
\begin{align}
K_{1}&=\dv{r}\dot{p_r}=\dv{r}\left(\frac{1}{\dot{t}}\frac{\partial \mathcal{L}}{\partial r}\right)\,,\\
K_{2}&=\dv{p_r}\dot{r}=\dv{p_r}\left(\frac{1}{\dot{t}}\frac{p_r}{q_{rr}}\right)=\frac{1}{\dot{t}q_{rr}}\,.
\end{align}

Since Lyapunov exponents are given by the eigenvalues of matrix $K_{ij}$, we need to diagonalize it. Applying the standard procedure, our characteristic equation takes the form
\begin{align}
\lambda^2=K_1 K_2=\frac{1}{\dot{t}q_{rr}}\dv{r}\left(\frac{1}{\dot{t}}\frac{\partial \mathcal{L}}{\partial r}\right)\,, \label{charlambda}
\end{align}
thence,
\begin{align}
\lambda=\sqrt{K_1K_2}\,, \label{lambda}
\end{align}
where we are only keeping the root with + sign as the principal Lyapunov exponent corresponds to the biggest eigenvalue.

All that is left is to calculate explicitly $K_1$ and $K_2$. To that end, we use the Euler--Lagrange equations of motion, in particular
\begin{align}
\dv{\mathcal{L}}{r}=\dv{\tau}\dv{\mathcal{L}}{\dot{r}}=\dv{\tau}(q_{rr}\dot{r})=\dot{r}\dv{r}(q_{rr}\dot{r})=\frac{1}{2q_{rr}}\dv{r}(q_{rr}^{2}\dot{r}^2)\,.
\end{align}
Using also the definition of the effective potential $V_r$, one gets
\begin{align}
\dv{\mathcal{L}}{r}=\frac{1}{2g_{rr}}\dv{r}(-g_{rr}^{2}V_r)\,.
\end{align}
From that and the fact that circular geodesics satisfy \eqref{circular geodesic condition 1} and \eqref{circular geodesic condition 2}, we finally obtain
\begin{align}
\lambda=\sqrt{-\frac{V_r^{''}}{2\dot{t}^2}}
\end{align}
which is exactly the formula  \eqref{def lambda} used in the main text.

\bibliography{Lyapunov.bib}

\begin{thebibliography}{53}%
\makeatletter
\providecommand \@ifxundefined [1]{%
 \@ifx{#1\undefined}
}%
\providecommand \@ifnum [1]{%
 \ifnum #1\expandafter \@firstoftwo
 \else \expandafter \@secondoftwo
 \fi
}%
\providecommand \@ifx [1]{%
 \ifx #1\expandafter \@firstoftwo
 \else \expandafter \@secondoftwo
 \fi
}%
\providecommand \natexlab [1]{#1}%
\providecommand \enquote  [1]{``#1''}%
\providecommand \bibnamefont  [1]{#1}%
\providecommand \bibfnamefont [1]{#1}%
\providecommand \citenamefont [1]{#1}%
\providecommand \href@noop [0]{\@secondoftwo}%
\providecommand \href [0]{\begingroup \@sanitize@url \@href}%
\providecommand \@href[1]{\@@startlink{#1}\@@href}%
\providecommand \@@href[1]{\endgroup#1\@@endlink}%
\providecommand \@sanitize@url [0]{\catcode `\\12\catcode `\$12\catcode
  `\&12\catcode `\#12\catcode `\^12\catcode `\_12\catcode `\%12\relax}%
\providecommand \@@startlink[1]{}%
\providecommand \@@endlink[0]{}%
\providecommand \url  [0]{\begingroup\@sanitize@url \@url }%
\providecommand \@url [1]{\endgroup\@href {#1}{\urlprefix }}%
\providecommand \urlprefix  [0]{URL }%
\providecommand \Eprint [0]{\href }%
\providecommand \doibase [0]{https://doi.org/}%
\providecommand \selectlanguage [0]{\@gobble}%
\providecommand \bibinfo  [0]{\@secondoftwo}%
\providecommand \bibfield  [0]{\@secondoftwo}%
\providecommand \translation [1]{[#1]}%
\providecommand \BibitemOpen [0]{}%
\providecommand \bibitemStop [0]{}%
\providecommand \bibitemNoStop [0]{.\EOS\space}%
\providecommand \EOS [0]{\spacefactor3000\relax}%
\providecommand \BibitemShut  [1]{\csname bibitem#1\endcsname}%
\let\auto@bib@innerbib\@empty
\bibitem [{\citenamefont {Born}(1933)}]{born1933modified}%
  \BibitemOpen
  \bibfield  {author} {\bibinfo {author} {\bibfnamefont {M.}~\bibnamefont
  {Born}},\ }\bibfield  {title} {\bibinfo {title} {Modified field equations
  with a finite radius of the electron},\ }\href
  {https://doi.org/10.1038/132282a0} {\bibfield  {journal} {\bibinfo  {journal}
  {Nature}\ }\textbf {\bibinfo {volume} {132}},\ \bibinfo {pages} {282}
  (\bibinfo {year} {1933})}\BibitemShut {NoStop}%
\bibitem [{\citenamefont {Born}\ and\ \citenamefont
  {Infeld}(1934{\natexlab{a}})}]{born1934foundations}%
  \BibitemOpen
  \bibfield  {author} {\bibinfo {author} {\bibfnamefont {M.}~\bibnamefont
  {Born}}\ and\ \bibinfo {author} {\bibfnamefont {L.}~\bibnamefont {Infeld}},\
  }\bibfield  {title} {\bibinfo {title} {Foundations of the new field theory},\
  }\href {https://doi.org/10.1098/rspa.1934.0059} {\bibfield  {journal}
  {\bibinfo  {journal} {Proceedings of the Royal Society of London. Series A,
  Containing Papers of a Mathematical and Physical Character}\ }\textbf
  {\bibinfo {volume} {144}},\ \bibinfo {pages} {425} (\bibinfo {year}
  {1934}{\natexlab{a}})}\BibitemShut {NoStop}%
\bibitem [{\citenamefont {Born}\ and\ \citenamefont
  {Infeld}(1934{\natexlab{b}})}]{born1934quantization}%
  \BibitemOpen
  \bibfield  {author} {\bibinfo {author} {\bibfnamefont {M.}~\bibnamefont
  {Born}}\ and\ \bibinfo {author} {\bibfnamefont {L.}~\bibnamefont {Infeld}},\
  }\bibfield  {title} {\bibinfo {title} {On the quantization of the new field
  equations {I}},\ }\href {https://doi.org/10.1098/rspa.1934.0234} {\bibfield
  {journal} {\bibinfo  {journal} {Proceedings of the Royal Society of London.
  Series A-Mathematical and Physical Sciences}\ }\textbf {\bibinfo {volume}
  {147}},\ \bibinfo {pages} {522} (\bibinfo {year}
  {1934}{\natexlab{b}})}\BibitemShut {NoStop}%
\bibitem [{\citenamefont {Hoffmann}\ and\ \citenamefont
  {Infeld}(1937)}]{HInfeld}%
  \BibitemOpen
  \bibfield  {author} {\bibinfo {author} {\bibfnamefont {B.}~\bibnamefont
  {Hoffmann}}\ and\ \bibinfo {author} {\bibfnamefont {L.}~\bibnamefont
  {Infeld}},\ }\bibfield  {title} {\bibinfo {title} {On the choice of the
  action function in the new field theory},\ }\href
  {https://doi.org/10.1103/PhysRev.51.765} {\bibfield  {journal} {\bibinfo
  {journal} {Phys. Rev.}\ }\textbf {\bibinfo {volume} {51}},\ \bibinfo {pages}
  {765} (\bibinfo {year} {1937})}\BibitemShut {NoStop}%
\bibitem [{\citenamefont {Sorokin}(2022)}]{Sorokin:2021tge}%
  \BibitemOpen
  \bibfield  {author} {\bibinfo {author} {\bibfnamefont {D.~P.}\ \bibnamefont
  {Sorokin}},\ }\bibfield  {title} {\bibinfo {title} {{Introductory Notes on
  Non-linear Electrodynamics and its Applications}},\ }\href
  {https://doi.org/10.1002/prop.202200092} {\bibfield  {journal} {\bibinfo
  {journal} {Fortsch. Phys.}\ }\textbf {\bibinfo {volume} {70}},\ \bibinfo
  {pages} {2200092} (\bibinfo {year} {2022})},\ \Eprint
  {https://arxiv.org/abs/2112.12118} {arXiv:2112.12118 [hep-th]} \BibitemShut
  {NoStop}%
\bibitem [{\citenamefont {Gibbons}\ and\ \citenamefont
  {Rasheed}(1995)}]{Gibbons:1995cv}%
  \BibitemOpen
  \bibfield  {author} {\bibinfo {author} {\bibfnamefont {G.~W.}\ \bibnamefont
  {Gibbons}}\ and\ \bibinfo {author} {\bibfnamefont {D.~A.}\ \bibnamefont
  {Rasheed}},\ }\bibfield  {title} {\bibinfo {title} {{Electric - magnetic
  duality rotations in nonlinear electrodynamics}},\ }\href
  {https://doi.org/10.1016/0550-3213(95)00409-L} {\bibfield  {journal}
  {\bibinfo  {journal} {Nucl. Phys. B}\ }\textbf {\bibinfo {volume} {454}},\
  \bibinfo {pages} {185} (\bibinfo {year} {1995})},\ \Eprint
  {https://arxiv.org/abs/hep-th/9506035} {arXiv:hep-th/9506035} \BibitemShut
  {NoStop}%
\bibitem [{\citenamefont {Fradkin}\ and\ \citenamefont
  {Tseytlin}(1985)}]{Fradkin:1985qd}%
  \BibitemOpen
  \bibfield  {author} {\bibinfo {author} {\bibfnamefont {E.~S.}\ \bibnamefont
  {Fradkin}}\ and\ \bibinfo {author} {\bibfnamefont {A.~A.}\ \bibnamefont
  {Tseytlin}},\ }\bibfield  {title} {\bibinfo {title} {{Nonlinear
  Electrodynamics from Quantized Strings}},\ }\href
  {https://doi.org/10.1016/0370-2693(85)90205-9} {\bibfield  {journal}
  {\bibinfo  {journal} {Phys. Lett. B}\ }\textbf {\bibinfo {volume} {163}},\
  \bibinfo {pages} {123} (\bibinfo {year} {1985})}\BibitemShut {NoStop}%
\bibitem [{\citenamefont {Leigh}(1989)}]{Leigh:1989jq}%
  \BibitemOpen
  \bibfield  {author} {\bibinfo {author} {\bibfnamefont {R.~G.}\ \bibnamefont
  {Leigh}},\ }\bibfield  {title} {\bibinfo {title} {{Dirac-Born-Infeld Action
  from Dirichlet Sigma Model}},\ }\href
  {https://doi.org/10.1142/S0217732389003099} {\bibfield  {journal} {\bibinfo
  {journal} {Mod. Phys. Lett. A}\ }\textbf {\bibinfo {volume} {4}},\ \bibinfo
  {pages} {2767} (\bibinfo {year} {1989})}\BibitemShut {NoStop}%
\bibitem [{\citenamefont {Callan}\ and\ \citenamefont
  {Maldacena}(1998)}]{Callan:1997kz}%
  \BibitemOpen
  \bibfield  {author} {\bibinfo {author} {\bibfnamefont {C.~G.}\ \bibnamefont
  {Callan}}\ and\ \bibinfo {author} {\bibfnamefont {J.~M.}\ \bibnamefont
  {Maldacena}},\ }\bibfield  {title} {\bibinfo {title} {{Brane death and
  dynamics from the Born-Infeld action}},\ }\href
  {https://doi.org/10.1016/S0550-3213(97)00700-1} {\bibfield  {journal}
  {\bibinfo  {journal} {Nucl. Phys. B}\ }\textbf {\bibinfo {volume} {513}},\
  \bibinfo {pages} {198} (\bibinfo {year} {1998})},\ \Eprint
  {https://arxiv.org/abs/hep-th/9708147} {arXiv:hep-th/9708147} \BibitemShut
  {NoStop}%
\bibitem [{\citenamefont {Alishahiha}\ \emph {et~al.}(2004)\citenamefont
  {Alishahiha}, \citenamefont {Silverstein},\ and\ \citenamefont
  {Tong}}]{Alishahiha:2004eh}%
  \BibitemOpen
  \bibfield  {author} {\bibinfo {author} {\bibfnamefont {M.}~\bibnamefont
  {Alishahiha}}, \bibinfo {author} {\bibfnamefont {E.}~\bibnamefont
  {Silverstein}},\ and\ \bibinfo {author} {\bibfnamefont {D.}~\bibnamefont
  {Tong}},\ }\bibfield  {title} {\bibinfo {title} {{DBI in the sky}},\ }\href
  {https://doi.org/10.1103/PhysRevD.70.123505} {\bibfield  {journal} {\bibinfo
  {journal} {Phys. Rev. D}\ }\textbf {\bibinfo {volume} {70}},\ \bibinfo
  {pages} {123505} (\bibinfo {year} {2004})},\ \Eprint
  {https://arxiv.org/abs/hep-th/0404084} {arXiv:hep-th/0404084} \BibitemShut
  {NoStop}%
\bibitem [{\citenamefont {Plebanski}(1970)}]{Plebanski:1970zz}%
  \BibitemOpen
  \bibfield  {author} {\bibinfo {author} {\bibfnamefont {J.}~\bibnamefont
  {Plebanski}},\ }\bibfield  {title} {\bibinfo {title} {{Lectures on non linear
  electrodynamics}},\ }\href@noop {} {\bibfield  {journal} {\bibinfo  {journal}
  {Copenhagen, Nordita}\ } (\bibinfo {year} {1970})}\BibitemShut {NoStop}%
\bibitem [{\citenamefont {Russo}\ and\ \citenamefont
  {Townsend}(2023)}]{Russo:2022qvz}%
  \BibitemOpen
  \bibfield  {author} {\bibinfo {author} {\bibfnamefont {J.~G.}\ \bibnamefont
  {Russo}}\ and\ \bibinfo {author} {\bibfnamefont {P.~K.}\ \bibnamefont
  {Townsend}},\ }\bibfield  {title} {\bibinfo {title} {{Nonlinear
  electrodynamics without birefringence}},\ }\href
  {https://doi.org/10.1007/JHEP01(2023)039} {\bibfield  {journal} {\bibinfo
  {journal} {JHEP}\ }\textbf {\bibinfo {volume} {01}},\ \bibinfo {pages}
  {039}},\ \Eprint {https://arxiv.org/abs/2211.10689} {arXiv:2211.10689
  [hep-th]} \BibitemShut {NoStop}%
\bibitem [{\citenamefont {Mezincescu}\ \emph {et~al.}(2023)\citenamefont
  {Mezincescu}, \citenamefont {Russo},\ and\ \citenamefont
  {Townsend}}]{Mezincescu:2023zny}%
  \BibitemOpen
  \bibfield  {author} {\bibinfo {author} {\bibfnamefont {L.}~\bibnamefont
  {Mezincescu}}, \bibinfo {author} {\bibfnamefont {J.~G.}\ \bibnamefont
  {Russo}},\ and\ \bibinfo {author} {\bibfnamefont {P.~K.}\ \bibnamefont
  {Townsend}},\ }\bibfield  {title} {\bibinfo {title} {{Hamiltonian
  birefringence and Born-Infeld limits}},\ }\href@noop {} {\  (\bibinfo {year}
  {2023})},\ \Eprint {https://arxiv.org/abs/2311.04278} {arXiv:2311.04278
  [hep-th]} \BibitemShut {NoStop}%
\bibitem [{\citenamefont {Novello}\ \emph {et~al.}(2000)\citenamefont
  {Novello}, \citenamefont {De~Lorenci}, \citenamefont {Salim},\ and\
  \citenamefont {Klippert}}]{Novello:1999pg}%
  \BibitemOpen
  \bibfield  {author} {\bibinfo {author} {\bibfnamefont {M.}~\bibnamefont
  {Novello}}, \bibinfo {author} {\bibfnamefont {V.~A.}\ \bibnamefont
  {De~Lorenci}}, \bibinfo {author} {\bibfnamefont {J.~M.}\ \bibnamefont
  {Salim}},\ and\ \bibinfo {author} {\bibfnamefont {R.}~\bibnamefont
  {Klippert}},\ }\bibfield  {title} {\bibinfo {title} {{Geometrical aspects of
  light propagation in nonlinear electrodynamics}},\ }\href
  {https://doi.org/10.1103/PhysRevD.61.045001} {\bibfield  {journal} {\bibinfo
  {journal} {Phys. Rev. D}\ }\textbf {\bibinfo {volume} {61}},\ \bibinfo
  {pages} {045001} (\bibinfo {year} {2000})},\ \Eprint
  {https://arxiv.org/abs/gr-qc/9911085} {arXiv:gr-qc/9911085} \BibitemShut
  {NoStop}%
\bibitem [{\citenamefont {Bardeen}(1968)}]{bardeen1968non}%
  \BibitemOpen
  \bibfield  {author} {\bibinfo {author} {\bibfnamefont {J.~M.}\ \bibnamefont
  {Bardeen}},\ }\bibfield  {title} {\bibinfo {title} {Non-singular
  general-relativistic gravitational collapse},\ }in\ \href@noop {} {\emph
  {\bibinfo {booktitle} {Proc. Int. Conf. GR5, Tbilisi}}},\ Vol.\ \bibinfo
  {volume} {174}\ (\bibinfo {year} {1968})\ p.\ \bibinfo {pages}
  {174}\BibitemShut {NoStop}%
\bibitem [{\citenamefont {Ayon-Beato}\ and\ \citenamefont
  {Garcia}(2000)}]{Bardeen}%
  \BibitemOpen
  \bibfield  {author} {\bibinfo {author} {\bibfnamefont {E.}~\bibnamefont
  {Ayon-Beato}}\ and\ \bibinfo {author} {\bibfnamefont {A.}~\bibnamefont
  {Garcia}},\ }\bibfield  {title} {\bibinfo {title} {{The Bardeen model as a
  nonlinear magnetic monopole}},\ }\href
  {https://doi.org/10.1016/S0370-2693(00)01125-4} {\bibfield  {journal}
  {\bibinfo  {journal} {Phys. Lett. B}\ }\textbf {\bibinfo {volume} {493}},\
  \bibinfo {pages} {149} (\bibinfo {year} {2000})},\ \Eprint
  {https://arxiv.org/abs/gr-qc/0009077} {arXiv:gr-qc/0009077} \BibitemShut
  {NoStop}%
\bibitem [{\citenamefont {Balart}\ and\ \citenamefont
  {Vagenas}(2014)}]{Balart:2014cga}%
  \BibitemOpen
  \bibfield  {author} {\bibinfo {author} {\bibfnamefont {L.}~\bibnamefont
  {Balart}}\ and\ \bibinfo {author} {\bibfnamefont {E.~C.}\ \bibnamefont
  {Vagenas}},\ }\bibfield  {title} {\bibinfo {title} {{Regular black holes with
  a nonlinear electrodynamics source}},\ }\href
  {https://doi.org/10.1103/PhysRevD.90.124045} {\bibfield  {journal} {\bibinfo
  {journal} {Phys. Rev. D}\ }\textbf {\bibinfo {volume} {90}},\ \bibinfo
  {pages} {124045} (\bibinfo {year} {2014})},\ \Eprint
  {https://arxiv.org/abs/1408.0306} {arXiv:1408.0306 [gr-qc]} \BibitemShut
  {NoStop}%
\bibitem [{\citenamefont {Fan}\ and\ \citenamefont {Wang}(2016)}]{Fan:2016hvf}%
  \BibitemOpen
  \bibfield  {author} {\bibinfo {author} {\bibfnamefont {Z.-Y.}\ \bibnamefont
  {Fan}}\ and\ \bibinfo {author} {\bibfnamefont {X.}~\bibnamefont {Wang}},\
  }\bibfield  {title} {\bibinfo {title} {{Construction of Regular Black Holes
  in General Relativity}},\ }\href {https://doi.org/10.1103/PhysRevD.94.124027}
  {\bibfield  {journal} {\bibinfo  {journal} {Phys. Rev. D}\ }\textbf {\bibinfo
  {volume} {94}},\ \bibinfo {pages} {124027} (\bibinfo {year} {2016})},\
  \Eprint {https://arxiv.org/abs/1610.02636} {arXiv:1610.02636 [gr-qc]}
  \BibitemShut {NoStop}%
\bibitem [{\citenamefont {Bokuli\'c}\ \emph {et~al.}(2023)\citenamefont
  {Bokuli\'c}, \citenamefont {Franzin}, \citenamefont {Juri\'c},\ and\
  \citenamefont {Smoli\'c}}]{Bokulic:2023afx}%
  \BibitemOpen
  \bibfield  {author} {\bibinfo {author} {\bibfnamefont {A.}~\bibnamefont
  {Bokuli\'c}}, \bibinfo {author} {\bibfnamefont {E.}~\bibnamefont {Franzin}},
  \bibinfo {author} {\bibfnamefont {T.}~\bibnamefont {Juri\'c}},\ and\ \bibinfo
  {author} {\bibfnamefont {I.}~\bibnamefont {Smoli\'c}},\ }\bibfield  {title}
  {\bibinfo {title} {{Lagrangian reverse engineering for regular black
  holes}},\ }\href@noop {} {\  (\bibinfo {year} {2023})},\ \Eprint
  {https://arxiv.org/abs/2311.17151} {arXiv:2311.17151 [gr-qc]} \BibitemShut
  {NoStop}%
\bibitem [{\citenamefont {Bandos}\ \emph {et~al.}(2020)\citenamefont {Bandos},
  \citenamefont {Lechner}, \citenamefont {Sorokin},\ and\ \citenamefont
  {Townsend}}]{ModMax}%
  \BibitemOpen
  \bibfield  {author} {\bibinfo {author} {\bibfnamefont {I.}~\bibnamefont
  {Bandos}}, \bibinfo {author} {\bibfnamefont {K.}~\bibnamefont {Lechner}},
  \bibinfo {author} {\bibfnamefont {D.}~\bibnamefont {Sorokin}},\ and\ \bibinfo
  {author} {\bibfnamefont {P.~K.}\ \bibnamefont {Townsend}},\ }\bibfield
  {title} {\bibinfo {title} {{Nonlinear duality-invariant conformal extension
  of {Maxwell’s} equations}},\ }\href
  {https://doi.org/10.1103/PhysRevD.102.121703} {\bibfield  {journal} {\bibinfo
   {journal} {Phys. Rev. D}\ }\textbf {\bibinfo {volume} {102}},\ \bibinfo
  {pages} {121703} (\bibinfo {year} {2020})}\BibitemShut {NoStop}%
\bibitem [{\citenamefont {Kosyakov}(2020)}]{Kosyakov:2020wxv}%
  \BibitemOpen
  \bibfield  {author} {\bibinfo {author} {\bibfnamefont {B.~P.}\ \bibnamefont
  {Kosyakov}},\ }\bibfield  {title} {\bibinfo {title} {{Nonlinear
  electrodynamics with the maximum allowable symmetries}},\ }\href
  {https://doi.org/10.1016/j.physletb.2020.135840} {\bibfield  {journal}
  {\bibinfo  {journal} {Phys. Lett. B}\ }\textbf {\bibinfo {volume} {810}},\
  \bibinfo {pages} {135840} (\bibinfo {year} {2020})},\ \Eprint
  {https://arxiv.org/abs/2007.13878} {arXiv:2007.13878 [hep-th]} \BibitemShut
  {NoStop}%
\bibitem [{\citenamefont {Russo}\ and\ \citenamefont
  {Townsend}(2024)}]{Russo:2024llm}%
  \BibitemOpen
  \bibfield  {author} {\bibinfo {author} {\bibfnamefont {J.~G.}\ \bibnamefont
  {Russo}}\ and\ \bibinfo {author} {\bibfnamefont {P.~K.}\ \bibnamefont
  {Townsend}},\ }\bibfield  {title} {\bibinfo {title} {{On Causal Self-Dual
  Electrodynamics}},\ }\href@noop {} {\  (\bibinfo {year} {2024})},\ \Eprint
  {https://arxiv.org/abs/2401.06707} {arXiv:2401.06707 [hep-th]} \BibitemShut
  {NoStop}%
\bibitem [{\citenamefont {Tahamtan}(2021)}]{newLagrangian}%
  \BibitemOpen
  \bibfield  {author} {\bibinfo {author} {\bibfnamefont {T.}~\bibnamefont
  {Tahamtan}},\ }\bibfield  {title} {\bibinfo {title} {{Compatibility of
  nonlinear electrodynamics models with Robinson-Trautman geometry}},\ }\href
  {https://doi.org/10.1103/PhysRevD.103.064052} {\bibfield  {journal} {\bibinfo
   {journal} {Phys. Rev. D}\ }\textbf {\bibinfo {volume} {103}},\ \bibinfo
  {pages} {064052} (\bibinfo {year} {2021})},\ \Eprint
  {https://arxiv.org/abs/2010.01689} {arXiv:2010.01689 [gr-qc]} \BibitemShut
  {NoStop}%
\bibitem [{\citenamefont {Tahamtan}\ \emph {et~al.}(2023)\citenamefont
  {Tahamtan}, \citenamefont {Flores-Alfonso},\ and\ \citenamefont
  {Svitek}}]{Tahamtan:2023tci}%
  \BibitemOpen
  \bibfield  {author} {\bibinfo {author} {\bibfnamefont {T.}~\bibnamefont
  {Tahamtan}}, \bibinfo {author} {\bibfnamefont {D.}~\bibnamefont
  {Flores-Alfonso}},\ and\ \bibinfo {author} {\bibfnamefont {O.}~\bibnamefont
  {Svitek}},\ }\bibfield  {title} {\bibinfo {title} {{Well-posed nonvacuum
  solutions in Robinson-Trautman geometry}},\ }\href
  {https://doi.org/10.1103/PhysRevD.108.124076} {\bibfield  {journal} {\bibinfo
   {journal} {Phys. Rev. D}\ }\textbf {\bibinfo {volume} {108}},\ \bibinfo
  {pages} {124076} (\bibinfo {year} {2023})},\ \Eprint
  {https://arxiv.org/abs/2311.03110} {arXiv:2311.03110 [gr-qc]} \BibitemShut
  {NoStop}%
\bibitem [{\citenamefont {Kubiznak}\ \emph {et~al.}(2022)\citenamefont
  {Kubiznak}, \citenamefont {Tahamtan},\ and\ \citenamefont
  {Svitek}}]{slowlyrotating}%
  \BibitemOpen
  \bibfield  {author} {\bibinfo {author} {\bibfnamefont {D.}~\bibnamefont
  {Kubiznak}}, \bibinfo {author} {\bibfnamefont {T.}~\bibnamefont {Tahamtan}},\
  and\ \bibinfo {author} {\bibfnamefont {O.}~\bibnamefont {Svitek}},\
  }\bibfield  {title} {\bibinfo {title} {{Slowly rotating black holes in
  nonlinear electrodynamics}},\ }\href
  {https://doi.org/10.1103/PhysRevD.105.104064} {\bibfield  {journal} {\bibinfo
   {journal} {Phys. Rev. D}\ }\textbf {\bibinfo {volume} {105}},\ \bibinfo
  {pages} {104064} (\bibinfo {year} {2022})},\ \Eprint
  {https://arxiv.org/abs/2203.01919} {arXiv:2203.01919 [gr-qc]} \BibitemShut
  {NoStop}%
\bibitem [{\citenamefont {Hale}\ \emph {et~al.}(2023)\citenamefont {Hale},
  \citenamefont {Kubiznak}, \citenamefont {Svitek},\ and\ \citenamefont
  {Tahamtan}}]{Kubiznak:2023emm}%
  \BibitemOpen
  \bibfield  {author} {\bibinfo {author} {\bibfnamefont {T.}~\bibnamefont
  {Hale}}, \bibinfo {author} {\bibfnamefont {D.}~\bibnamefont {Kubiznak}},
  \bibinfo {author} {\bibfnamefont {O.}~\bibnamefont {Svitek}},\ and\ \bibinfo
  {author} {\bibfnamefont {T.}~\bibnamefont {Tahamtan}},\ }\bibfield  {title}
  {\bibinfo {title} {{Solutions and basic properties of regularized
  \uppercase{M}axwell theory}},\ }\href
  {https://doi.org/10.1103/PhysRevD.107.124031} {\bibfield  {journal} {\bibinfo
   {journal} {Phys. Rev. D}\ }\textbf {\bibinfo {volume} {107}},\ \bibinfo
  {pages} {124031} (\bibinfo {year} {2023})},\ \Eprint
  {https://arxiv.org/abs/2303.16928} {arXiv:2303.16928 [gr-qc]} \BibitemShut
  {NoStop}%
\bibitem [{\citenamefont {Zhang}\ \emph {et~al.}(2019)\citenamefont {Zhang},
  \citenamefont {Han}, \citenamefont {Jiang},\ and\ \citenamefont
  {Liu}}]{Zhang:2019tzi}%
  \BibitemOpen
  \bibfield  {author} {\bibinfo {author} {\bibfnamefont {M.}~\bibnamefont
  {Zhang}}, \bibinfo {author} {\bibfnamefont {S.-Z.}\ \bibnamefont {Han}},
  \bibinfo {author} {\bibfnamefont {J.}~\bibnamefont {Jiang}},\ and\ \bibinfo
  {author} {\bibfnamefont {W.-B.}\ \bibnamefont {Liu}},\ }\bibfield  {title}
  {\bibinfo {title} {{Circular orbit of a test particle and phase transition of
  a black hole}},\ }\href {https://doi.org/10.1103/PhysRevD.99.065016}
  {\bibfield  {journal} {\bibinfo  {journal} {Phys. Rev. D}\ }\textbf {\bibinfo
  {volume} {99}},\ \bibinfo {pages} {065016} (\bibinfo {year} {2019})},\
  \Eprint {https://arxiv.org/abs/1903.08293} {arXiv:1903.08293 [hep-th]}
  \BibitemShut {NoStop}%
\bibitem [{\citenamefont {Guo}\ \emph {et~al.}(2022)\citenamefont {Guo},
  \citenamefont {Lu}, \citenamefont {Mu},\ and\ \citenamefont
  {Wang}}]{Guo:2022kio}%
  \BibitemOpen
  \bibfield  {author} {\bibinfo {author} {\bibfnamefont {X.}~\bibnamefont
  {Guo}}, \bibinfo {author} {\bibfnamefont {Y.}~\bibnamefont {Lu}}, \bibinfo
  {author} {\bibfnamefont {B.}~\bibnamefont {Mu}},\ and\ \bibinfo {author}
  {\bibfnamefont {P.}~\bibnamefont {Wang}},\ }\bibfield  {title} {\bibinfo
  {title} {{Probing phase structure of black holes with Lyapunov exponents}},\
  }\href {https://doi.org/10.1007/JHEP08(2022)153} {\bibfield  {journal}
  {\bibinfo  {journal} {JHEP}\ }\textbf {\bibinfo {volume} {08}},\ \bibinfo
  {pages} {153}},\ \Eprint {https://arxiv.org/abs/2205.02122} {arXiv:2205.02122
  [gr-qc]} \BibitemShut {NoStop}%
\bibitem [{\citenamefont {Yang}\ \emph {et~al.}(2023)\citenamefont {Yang},
  \citenamefont {Tao}, \citenamefont {Mu},\ and\ \citenamefont
  {He}}]{Yang:2023hci}%
  \BibitemOpen
  \bibfield  {author} {\bibinfo {author} {\bibfnamefont {S.}~\bibnamefont
  {Yang}}, \bibinfo {author} {\bibfnamefont {J.}~\bibnamefont {Tao}}, \bibinfo
  {author} {\bibfnamefont {B.}~\bibnamefont {Mu}},\ and\ \bibinfo {author}
  {\bibfnamefont {A.}~\bibnamefont {He}},\ }\bibfield  {title} {\bibinfo
  {title} {{Lyapunov Exponents and Phase Transitions of Born-Infeld AdS Black
  Holes}},\ }\href@noop {} {\  (\bibinfo {year} {2023})},\ \Eprint
  {https://arxiv.org/abs/2304.01877} {arXiv:2304.01877 [gr-qc]} \BibitemShut
  {NoStop}%
\bibitem [{\citenamefont {Lyu}\ \emph {et~al.}(2023)\citenamefont {Lyu},
  \citenamefont {Tao},\ and\ \citenamefont {Wang}}]{Lyu:2023sih}%
  \BibitemOpen
  \bibfield  {author} {\bibinfo {author} {\bibfnamefont {X.}~\bibnamefont
  {Lyu}}, \bibinfo {author} {\bibfnamefont {J.}~\bibnamefont {Tao}},\ and\
  \bibinfo {author} {\bibfnamefont {P.}~\bibnamefont {Wang}},\ }\bibfield
  {title} {\bibinfo {title} {{Probing the thermodynamics of charged Gauss
  Bonnet AdS black holes with the Lyapunov exponent}},\ }\href@noop {} {\
  (\bibinfo {year} {2023})},\ \Eprint {https://arxiv.org/abs/2312.11912}
  {arXiv:2312.11912 [gr-qc]} \BibitemShut {NoStop}%
\bibitem [{\citenamefont {Kumara}\ \emph {et~al.}(2024)\citenamefont {Kumara},
  \citenamefont {Punacha},\ and\ \citenamefont {Ali}}]{Kumara:2024obd}%
  \BibitemOpen
  \bibfield  {author} {\bibinfo {author} {\bibfnamefont {A.~N.}\ \bibnamefont
  {Kumara}}, \bibinfo {author} {\bibfnamefont {S.}~\bibnamefont {Punacha}},\
  and\ \bibinfo {author} {\bibfnamefont {M.~S.}\ \bibnamefont {Ali}},\
  }\bibfield  {title} {\bibinfo {title} {{Lyapunov Exponents and Phase
  Structure of Lifshitz and Hyperscaling Violating Black Holes}},\ }\href@noop
  {} {\  (\bibinfo {year} {2024})},\ \Eprint {https://arxiv.org/abs/2401.05181}
  {arXiv:2401.05181 [gr-qc]} \BibitemShut {NoStop}%
\bibitem [{\citenamefont {Cai}\ and\ \citenamefont {Miao}(2021)}]{Cai:2021uov}%
  \BibitemOpen
  \bibfield  {author} {\bibinfo {author} {\bibfnamefont {X.-C.}\ \bibnamefont
  {Cai}}\ and\ \bibinfo {author} {\bibfnamefont {Y.-G.}\ \bibnamefont {Miao}},\
  }\bibfield  {title} {\bibinfo {title} {Can we know about black hole
  thermodynamics through shadows?},\ }\href@noop {} {\  (\bibinfo {year}
  {2021})},\ \Eprint {https://arxiv.org/abs/2107.08352} {arXiv:2107.08352
  [gr-qc]} \BibitemShut {NoStop}%
\bibitem [{\citenamefont {Cardoso}\ \emph {et~al.}(2009)\citenamefont
  {Cardoso}, \citenamefont {Miranda}, \citenamefont {Berti}, \citenamefont
  {Witek},\ and\ \citenamefont {Zanchin}}]{Cardoso:2008bp}%
  \BibitemOpen
  \bibfield  {author} {\bibinfo {author} {\bibfnamefont {V.}~\bibnamefont
  {Cardoso}}, \bibinfo {author} {\bibfnamefont {A.~S.}\ \bibnamefont
  {Miranda}}, \bibinfo {author} {\bibfnamefont {E.}~\bibnamefont {Berti}},
  \bibinfo {author} {\bibfnamefont {H.}~\bibnamefont {Witek}},\ and\ \bibinfo
  {author} {\bibfnamefont {V.~T.}\ \bibnamefont {Zanchin}},\ }\bibfield
  {title} {\bibinfo {title} {{Geodesic stability, Lyapunov exponents and
  quasinormal modes}},\ }\href {https://doi.org/10.1103/PhysRevD.79.064016}
  {\bibfield  {journal} {\bibinfo  {journal} {Phys. Rev. D}\ }\textbf {\bibinfo
  {volume} {79}},\ \bibinfo {pages} {064016} (\bibinfo {year} {2009})},\
  \Eprint {https://arxiv.org/abs/0812.1806} {arXiv:0812.1806 [hep-th]}
  \BibitemShut {NoStop}%
\bibitem [{\citenamefont {Hadar}\ \emph {et~al.}(2022)\citenamefont {Hadar},
  \citenamefont {Kapec}, \citenamefont {Lupsasca},\ and\ \citenamefont
  {Strominger}}]{Hadar:2022xag}%
  \BibitemOpen
  \bibfield  {author} {\bibinfo {author} {\bibfnamefont {S.}~\bibnamefont
  {Hadar}}, \bibinfo {author} {\bibfnamefont {D.}~\bibnamefont {Kapec}},
  \bibinfo {author} {\bibfnamefont {A.}~\bibnamefont {Lupsasca}},\ and\
  \bibinfo {author} {\bibfnamefont {A.}~\bibnamefont {Strominger}},\ }\bibfield
   {title} {\bibinfo {title} {{Holography of the photon ring}},\ }\href
  {https://doi.org/10.1088/1361-6382/ac8d43} {\bibfield  {journal} {\bibinfo
  {journal} {Class. Quant. Grav.}\ }\textbf {\bibinfo {volume} {39}},\ \bibinfo
  {pages} {215001} (\bibinfo {year} {2022})},\ \Eprint
  {https://arxiv.org/abs/2205.05064} {arXiv:2205.05064 [gr-qc]} \BibitemShut
  {NoStop}%
\bibitem [{\citenamefont {Kapec}\ \emph {et~al.}(2023)\citenamefont {Kapec},
  \citenamefont {Lupsasca},\ and\ \citenamefont {Strominger}}]{Kapec:2022dvc}%
  \BibitemOpen
  \bibfield  {author} {\bibinfo {author} {\bibfnamefont {D.}~\bibnamefont
  {Kapec}}, \bibinfo {author} {\bibfnamefont {A.}~\bibnamefont {Lupsasca}},\
  and\ \bibinfo {author} {\bibfnamefont {A.}~\bibnamefont {Strominger}},\
  }\bibfield  {title} {\bibinfo {title} {{Photon rings around warped black
  holes}},\ }\href {https://doi.org/10.1088/1361-6382/acc164} {\bibfield
  {journal} {\bibinfo  {journal} {Class. Quant. Grav.}\ }\textbf {\bibinfo
  {volume} {40}},\ \bibinfo {pages} {095006} (\bibinfo {year} {2023})},\
  \Eprint {https://arxiv.org/abs/2211.01674} {arXiv:2211.01674 [gr-qc]}
  \BibitemShut {NoStop}%
\bibitem [{\citenamefont {Johnson}\ \emph {et~al.}(2020)\citenamefont {Johnson}
  \emph {et~al.}}]{Johnson:2019ljv}%
  \BibitemOpen
  \bibfield  {author} {\bibinfo {author} {\bibfnamefont {M.~D.}\ \bibnamefont
  {Johnson}} \emph {et~al.},\ }\bibfield  {title} {\bibinfo {title} {{Universal
  interferometric signatures of a black hole\textquoteright{}s photon ring}},\
  }\href {https://doi.org/10.1126/sciadv.aaz1310} {\bibfield  {journal}
  {\bibinfo  {journal} {Sci. Adv.}\ }\textbf {\bibinfo {volume} {6}},\ \bibinfo
  {pages} {eaaz1310} (\bibinfo {year} {2020})},\ \Eprint
  {https://arxiv.org/abs/1907.04329} {arXiv:1907.04329 [astro-ph.IM]}
  \BibitemShut {NoStop}%
\bibitem [{\citenamefont {Gralla}\ \emph {et~al.}(2020)\citenamefont {Gralla},
  \citenamefont {Lupsasca},\ and\ \citenamefont {Marrone}}]{Gralla:2020srx}%
  \BibitemOpen
  \bibfield  {author} {\bibinfo {author} {\bibfnamefont {S.~E.}\ \bibnamefont
  {Gralla}}, \bibinfo {author} {\bibfnamefont {A.}~\bibnamefont {Lupsasca}},\
  and\ \bibinfo {author} {\bibfnamefont {D.~P.}\ \bibnamefont {Marrone}},\
  }\bibfield  {title} {\bibinfo {title} {{The shape of the black hole photon
  ring: A precise test of strong-field general relativity}},\ }\href
  {https://doi.org/10.1103/PhysRevD.102.124004} {\bibfield  {journal} {\bibinfo
   {journal} {Phys. Rev. D}\ }\textbf {\bibinfo {volume} {102}},\ \bibinfo
  {pages} {124004} (\bibinfo {year} {2020})},\ \Eprint
  {https://arxiv.org/abs/2008.03879} {arXiv:2008.03879 [gr-qc]} \BibitemShut
  {NoStop}%
\bibitem [{\citenamefont {Wald}(1984)}]{Wald:1984rg}%
  \BibitemOpen
  \bibfield  {author} {\bibinfo {author} {\bibfnamefont {R.~M.}\ \bibnamefont
  {Wald}},\ }\href {https://doi.org/10.7208/chicago/9780226870373.001.0001}
  {\emph {\bibinfo {title} {{General Relativity}}}}\ (\bibinfo  {publisher}
  {Chicago Univ. Pr.},\ \bibinfo {address} {Chicago, USA},\ \bibinfo {year}
  {1984})\BibitemShut {NoStop}%
\bibitem [{\citenamefont {Kastor}\ \emph {et~al.}(2009)\citenamefont {Kastor},
  \citenamefont {Ray},\ and\ \citenamefont {Traschen}}]{Kastor:2009wy}%
  \BibitemOpen
  \bibfield  {author} {\bibinfo {author} {\bibfnamefont {D.}~\bibnamefont
  {Kastor}}, \bibinfo {author} {\bibfnamefont {S.}~\bibnamefont {Ray}},\ and\
  \bibinfo {author} {\bibfnamefont {J.}~\bibnamefont {Traschen}},\ }\bibfield
  {title} {\bibinfo {title} {{Enthalpy and the Mechanics of AdS Black Holes}},\
  }\href {https://doi.org/10.1088/0264-9381/26/19/195011} {\bibfield  {journal}
  {\bibinfo  {journal} {Class. Quant. Grav.}\ }\textbf {\bibinfo {volume}
  {26}},\ \bibinfo {pages} {195011} (\bibinfo {year} {2009})},\ \Eprint
  {https://arxiv.org/abs/0904.2765} {arXiv:0904.2765 [hep-th]} \BibitemShut
  {NoStop}%
\bibitem [{\citenamefont {Kubiznak}\ \emph {et~al.}(2017)\citenamefont
  {Kubiznak}, \citenamefont {Mann},\ and\ \citenamefont
  {Teo}}]{Kubiznak:2016qmn}%
  \BibitemOpen
  \bibfield  {author} {\bibinfo {author} {\bibfnamefont {D.}~\bibnamefont
  {Kubiznak}}, \bibinfo {author} {\bibfnamefont {R.~B.}\ \bibnamefont {Mann}},\
  and\ \bibinfo {author} {\bibfnamefont {M.}~\bibnamefont {Teo}},\ }\bibfield
  {title} {\bibinfo {title} {{Black hole chemistry: thermodynamics with
  Lambda}},\ }\href {https://doi.org/10.1088/1361-6382/aa5c69} {\bibfield
  {journal} {\bibinfo  {journal} {Class. Quant. Grav.}\ }\textbf {\bibinfo
  {volume} {34}},\ \bibinfo {pages} {063001} (\bibinfo {year} {2017})},\
  \Eprint {https://arxiv.org/abs/1608.06147} {arXiv:1608.06147 [hep-th]}
  \BibitemShut {NoStop}%
\bibitem [{\citenamefont {Gunasekaran}\ \emph {et~al.}(2012)\citenamefont
  {Gunasekaran}, \citenamefont {Mann},\ and\ \citenamefont
  {Kubiznak}}]{Gunasekaran:2012dq}%
  \BibitemOpen
  \bibfield  {author} {\bibinfo {author} {\bibfnamefont {S.}~\bibnamefont
  {Gunasekaran}}, \bibinfo {author} {\bibfnamefont {R.~B.}\ \bibnamefont
  {Mann}},\ and\ \bibinfo {author} {\bibfnamefont {D.}~\bibnamefont
  {Kubiznak}},\ }\bibfield  {title} {\bibinfo {title} {{Extended phase space
  thermodynamics for charged and rotating black holes and Born-Infeld vacuum
  polarization}},\ }\href {https://doi.org/10.1007/JHEP11(2012)110} {\bibfield
  {journal} {\bibinfo  {journal} {JHEP}\ }\textbf {\bibinfo {volume} {11}},\
  \bibinfo {pages} {110}},\ \Eprint {https://arxiv.org/abs/1208.6251}
  {arXiv:1208.6251 [hep-th]} \BibitemShut {NoStop}%
\bibitem [{\citenamefont {Chen}\ \emph {et~al.}(2023)\citenamefont {Chen},
  \citenamefont {Wang},\ and\ \citenamefont {Yang}}]{chen2023interferometric}%
  \BibitemOpen
  \bibfield  {author} {\bibinfo {author} {\bibfnamefont {Y.}~\bibnamefont
  {Chen}}, \bibinfo {author} {\bibfnamefont {P.}~\bibnamefont {Wang}},\ and\
  \bibinfo {author} {\bibfnamefont {H.}~\bibnamefont {Yang}},\ }\href@noop {}
  {\bibinfo {title} {Interferometric signatures of black holes with multiple
  photon spheres}} (\bibinfo {year} {2023}),\ \Eprint
  {https://arxiv.org/abs/2312.10304} {arXiv:2312.10304 [gr-qc]} \BibitemShut
  {NoStop}%
\bibitem [{\citenamefont {Pradhan}\ and\ \citenamefont
  {Majumdar}(2011)}]{Pradhan:2010ws}%
  \BibitemOpen
  \bibfield  {author} {\bibinfo {author} {\bibfnamefont {P.}~\bibnamefont
  {Pradhan}}\ and\ \bibinfo {author} {\bibfnamefont {P.}~\bibnamefont
  {Majumdar}},\ }\bibfield  {title} {\bibinfo {title} {{Circular Orbits in
  Extremal Reissner Nordstrom Spacetimes}},\ }\href
  {https://doi.org/10.1016/j.physleta.2010.11.015} {\bibfield  {journal}
  {\bibinfo  {journal} {Phys. Lett. A}\ }\textbf {\bibinfo {volume} {375}},\
  \bibinfo {pages} {474} (\bibinfo {year} {2011})},\ \Eprint
  {https://arxiv.org/abs/1001.0359} {arXiv:1001.0359 [gr-qc]} \BibitemShut
  {NoStop}%
\bibitem [{\citenamefont {Khoo}\ and\ \citenamefont
  {Ong}(2016)}]{Khoo:2016xqv}%
  \BibitemOpen
  \bibfield  {author} {\bibinfo {author} {\bibfnamefont {F.~S.}\ \bibnamefont
  {Khoo}}\ and\ \bibinfo {author} {\bibfnamefont {Y.~C.}\ \bibnamefont {Ong}},\
  }\bibfield  {title} {\bibinfo {title} {{Lux in obscuro: Photon Orbits of
  Extremal Black Holes Revisited}},\ }\href
  {https://doi.org/10.1088/0264-9381/33/23/235002} {\bibfield  {journal}
  {\bibinfo  {journal} {Class. Quant. Grav.}\ }\textbf {\bibinfo {volume}
  {33}},\ \bibinfo {pages} {235002} (\bibinfo {year} {2016})},\ \bibinfo {note}
  {[Erratum: Class.Quant.Grav. 34, 219501 (2017)]},\ \Eprint
  {https://arxiv.org/abs/1605.05774} {arXiv:1605.05774 [gr-qc]} \BibitemShut
  {NoStop}%
\bibitem [{\citenamefont {Liberati}\ \emph {et~al.}(2001)\citenamefont
  {Liberati}, \citenamefont {Sonego},\ and\ \citenamefont
  {Visser}}]{Liberati:2000mp}%
  \BibitemOpen
  \bibfield  {author} {\bibinfo {author} {\bibfnamefont {S.}~\bibnamefont
  {Liberati}}, \bibinfo {author} {\bibfnamefont {S.}~\bibnamefont {Sonego}},\
  and\ \bibinfo {author} {\bibfnamefont {M.}~\bibnamefont {Visser}},\
  }\bibfield  {title} {\bibinfo {title} {{Scharnhorst effect at oblique
  incidence}},\ }\href {https://doi.org/10.1103/PhysRevD.63.085003} {\bibfield
  {journal} {\bibinfo  {journal} {Phys. Rev. D}\ }\textbf {\bibinfo {volume}
  {63}},\ \bibinfo {pages} {085003} (\bibinfo {year} {2001})},\ \Eprint
  {https://arxiv.org/abs/quant-ph/0010055} {arXiv:quant-ph/0010055}
  \BibitemShut {NoStop}%
\bibitem [{\citenamefont {Liberati}\ \emph {et~al.}(2002)\citenamefont
  {Liberati}, \citenamefont {Sonego},\ and\ \citenamefont
  {Visser}}]{Liberati:2001sd}%
  \BibitemOpen
  \bibfield  {author} {\bibinfo {author} {\bibfnamefont {S.}~\bibnamefont
  {Liberati}}, \bibinfo {author} {\bibfnamefont {S.}~\bibnamefont {Sonego}},\
  and\ \bibinfo {author} {\bibfnamefont {M.}~\bibnamefont {Visser}},\
  }\bibfield  {title} {\bibinfo {title} {{Faster than c signals, special
  relativity, and causality}},\ }\href {https://doi.org/10.1006/aphy.2002.6233}
  {\bibfield  {journal} {\bibinfo  {journal} {Annals Phys.}\ }\textbf {\bibinfo
  {volume} {298}},\ \bibinfo {pages} {167} (\bibinfo {year} {2002})},\ \Eprint
  {https://arxiv.org/abs/gr-qc/0107091} {arXiv:gr-qc/0107091} \BibitemShut
  {NoStop}%
\bibitem [{\citenamefont {Shore}(2002)}]{Shore:2002gn}%
  \BibitemOpen
  \bibfield  {author} {\bibinfo {author} {\bibfnamefont {G.~M.}\ \bibnamefont
  {Shore}},\ }\bibfield  {title} {\bibinfo {title} {{Faster than light photons
  in gravitational fields. 2. Dispersion and vacuum polarization}},\ }\href
  {https://doi.org/10.1016/S0550-3213(02)00240-7} {\bibfield  {journal}
  {\bibinfo  {journal} {Nucl. Phys. B}\ }\textbf {\bibinfo {volume} {633}},\
  \bibinfo {pages} {271} (\bibinfo {year} {2002})},\ \Eprint
  {https://arxiv.org/abs/gr-qc/0203034} {arXiv:gr-qc/0203034} \BibitemShut
  {NoStop}%
\bibitem [{\citenamefont {Hajian}\ \emph {et~al.}(2021)\citenamefont {Hajian},
  \citenamefont {Liberati}, \citenamefont {Sheikh-Jabbari},\ and\ \citenamefont
  {Vahidinia}}]{Hajian:2020dcq}%
  \BibitemOpen
  \bibfield  {author} {\bibinfo {author} {\bibfnamefont {K.}~\bibnamefont
  {Hajian}}, \bibinfo {author} {\bibfnamefont {S.}~\bibnamefont {Liberati}},
  \bibinfo {author} {\bibfnamefont {M.~M.}\ \bibnamefont {Sheikh-Jabbari}},\
  and\ \bibinfo {author} {\bibfnamefont {M.~H.}\ \bibnamefont {Vahidinia}},\
  }\bibfield  {title} {\bibinfo {title} {{On Black Hole Temperature in
  Horndeski Gravity}},\ }\href {https://doi.org/10.1016/j.physletb.2020.136002}
  {\bibfield  {journal} {\bibinfo  {journal} {Phys. Lett. B}\ }\textbf
  {\bibinfo {volume} {812}},\ \bibinfo {pages} {136002} (\bibinfo {year}
  {2021})},\ \Eprint {https://arxiv.org/abs/2005.12985} {arXiv:2005.12985
  [gr-qc]} \BibitemShut {NoStop}%
\bibitem [{\citenamefont {Perlick}\ and\ \citenamefont
  {Tsupko}(2022)}]{Perlick:2021aok}%
  \BibitemOpen
  \bibfield  {author} {\bibinfo {author} {\bibfnamefont {V.}~\bibnamefont
  {Perlick}}\ and\ \bibinfo {author} {\bibfnamefont {O.~Y.}\ \bibnamefont
  {Tsupko}},\ }\bibfield  {title} {\bibinfo {title} {{Calculating black hole
  shadows: Review of analytical studies}},\ }\href
  {https://doi.org/10.1016/j.physrep.2021.10.004} {\bibfield  {journal}
  {\bibinfo  {journal} {Phys. Rept.}\ }\textbf {\bibinfo {volume} {947}},\
  \bibinfo {pages} {1} (\bibinfo {year} {2022})},\ \Eprint
  {https://arxiv.org/abs/2105.07101} {arXiv:2105.07101 [gr-qc]} \BibitemShut
  {NoStop}%
\bibitem [{\citenamefont {Kubiznak}\ and\ \citenamefont
  {Mann}(2012)}]{Kubiznak:2012wp}%
  \BibitemOpen
  \bibfield  {author} {\bibinfo {author} {\bibfnamefont {D.}~\bibnamefont
  {Kubiznak}}\ and\ \bibinfo {author} {\bibfnamefont {R.~B.}\ \bibnamefont
  {Mann}},\ }\bibfield  {title} {\bibinfo {title} {{P-V criticality of charged
  AdS black holes}},\ }\href {https://doi.org/10.1007/JHEP07(2012)033}
  {\bibfield  {journal} {\bibinfo  {journal} {JHEP}\ }\textbf {\bibinfo
  {volume} {07}},\ \bibinfo {pages} {033}},\ \Eprint
  {https://arxiv.org/abs/1205.0559} {arXiv:1205.0559 [hep-th]} \BibitemShut
  {NoStop}%
\bibitem [{\citenamefont {Chamblin}\ \emph {et~al.}(1999)\citenamefont
  {Chamblin}, \citenamefont {Emparan}, \citenamefont {Johnson},\ and\
  \citenamefont {Myers}}]{Chamblin:1999tk}%
  \BibitemOpen
  \bibfield  {author} {\bibinfo {author} {\bibfnamefont {A.}~\bibnamefont
  {Chamblin}}, \bibinfo {author} {\bibfnamefont {R.}~\bibnamefont {Emparan}},
  \bibinfo {author} {\bibfnamefont {C.~V.}\ \bibnamefont {Johnson}},\ and\
  \bibinfo {author} {\bibfnamefont {R.~C.}\ \bibnamefont {Myers}},\ }\bibfield
  {title} {\bibinfo {title} {{Charged AdS black holes and catastrophic
  holography}},\ }\href {https://doi.org/10.1103/PhysRevD.60.064018} {\bibfield
   {journal} {\bibinfo  {journal} {Phys. Rev. D}\ }\textbf {\bibinfo {volume}
  {60}},\ \bibinfo {pages} {064018} (\bibinfo {year} {1999})},\ \Eprint
  {https://arxiv.org/abs/hep-th/9902170} {arXiv:hep-th/9902170} \BibitemShut
  {NoStop}%
\bibitem [{\citenamefont {Giri}\ and\ \citenamefont
  {Nandan}(2021)}]{Springerlambda}%
  \BibitemOpen
  \bibfield  {author} {\bibinfo {author} {\bibfnamefont {S.}~\bibnamefont
  {Giri}}\ and\ \bibinfo {author} {\bibfnamefont {H.}~\bibnamefont {Nandan}},\
  }\bibfield  {title} {\bibinfo {title} {{Stability analysis of geodesics and
  quasinormal modes of a dual stringy black hole via Lyapunov exponents}},\
  }\href {https://doi.org/10.1007/s10714-021-02845-9} {\bibfield  {journal}
  {\bibinfo  {journal} {Gen. Relativ. Gravit.}\ }\textbf {\bibinfo {volume}
  {53}},\ \bibinfo {pages} {76} (\bibinfo {year} {2021})},\ \Eprint
  {https://arxiv.org/abs/2108.05772} {arXiv:2108.05772 [gr-qc]} \BibitemShut
  {NoStop}%
\bibitem [{\citenamefont {Pradhan}(2013)}]{Springerlambda2}%
  \BibitemOpen
  \bibfield  {author} {\bibinfo {author} {\bibfnamefont {P.}~\bibnamefont
  {Pradhan}},\ }\bibfield  {title} {\bibinfo {title} {{Lyapunov exponent and
  charged Myers–Perry spacetimes}},\ }\href
  {https://doi.org/10.1140/epjc/s10052-013-2477-8} {\bibfield  {journal}
  {\bibinfo  {journal} {Eur. Phys. J. C}\ }\textbf {\bibinfo {volume} {73}},\
  \bibinfo {pages} {2477} (\bibinfo {year} {2013})},\ \Eprint
  {https://arxiv.org/abs/1302.2536} {arXiv:1302.2536 [gr-qc]} \BibitemShut
  {NoStop}%
\end{thebibliography}%

\end{document}